\documentclass[preprint]{aastex6}
\usepackage{graphicx} 
\usepackage{natbib} 
\usepackage{amsmath} 
\usepackage{multirow}
\bibpunct{(}{)}{;}{a}{}{,}

\newcommand{\mean}[1]{\mbox{$\langle#1\rangle$}} 
\newcommand{\jj}[2]{\mbox{$J = #1\rightarrow#2$}}
\newcommand{\lsun}{\mbox{L$_\odot$}}
\newcommand{\msun}{\mbox{M$_\odot$}}
\newcommand{\msunmyr}{\mbox{M$_\odot$} Myr$^{-1}$}
\newcommand{\ee}[1]{\mbox{${} \times 10^{#1}$}}
\newcommand{\eten}[1]{\mbox{$10^{#1}$}}

\newcommand{\sfrmir}{\mbox{\rm SFR(MIR)}}

\newcommand{\mic}{\mbox{$\mu$m}}

\newcommand{\ssfr}{\mbox{$\Sigma_{\text{SFR}}$}}
\newcommand{\tdep}{\mbox{$t_{\rm{dep}}$}}
\newcommand{\co}{\textsuperscript{13}CO}
\newcommand{\coo}{\mbox{$^{13}$CO}}
\newcommand{\cotw}{\textsuperscript{12}CO}
\newcommand{\vlsr}{\mbox{$v_{LSR}$}}
\newcommand{\vrrl}{\mbox{$v_{\rm RRL}$}}
\newcommand{\kms}{km s$^{-1}$}

\newcommand{\md}{\mbox{$M_{\text{dense}}$}}
\newcommand{\tff}{\mbox{$t_{\rm ff}$}}
\newcommand{\tdyn}{\mbox{$t_{\rm dyn}$}}
\newcommand{\epsff}{\mbox{$\epsilon_{\rm ff}$}}
\newcommand{\hh}{\mbox{{\rm H}$_2$}}
\newcommand{\hcop}{\mbox{{\rm HCO}$^+$}}
\newcommand{\nthp}{\mbox{{\rm N}$_2${\rm H}$^+$}}
\newcommand{\hii}{\mbox{\ion{H}{2}}}
\newcommand{\mvir}{\mbox{$M_{\rm vir}$}}
\newcommand{\mcloud}{\mbox{$M_{\rm cloud}$}}
\newcommand{\mmol}{\mbox{$M_{\rm mol}$}}
\newcommand{\rcloud}{\mbox{$r_{\rm cloud}$}}
\newcommand{\ncloud}{\mbox{$n_{\rm cloud}$}}
\newcommand{\mdense}{\mbox{$M_{\rm dense}$}}
\newcommand{\rdense}{\mbox{$r_{\rm dense}$}}
\newcommand{\sfr}{\mbox{\rm SFR}}
\newcommand{\sfe}{\mbox{\rm SFE}}
\newcommand{\av}{\mbox{$A_V$}}
\newcommand\cmv{\mbox{cm$^{-3}$}}

\newcommand\gmc{\mbox{MC}}
\newcommand\ulirg{\mbox{(U)LIRG}}
\newcommand{\tex}{\mbox{$T_{\rm ex}$}}
\newcommand{\ammonia}{\mbox{{\rm NH}$_3$}}
\newcommand{\mclp}{\mbox{$m_{\rm cloud}$}}
\newcommand{\thetacl}{\mbox{$\theta_{\rm cloud}$}}
\newcommand{\sfrp}{\mbox{sfr}}

\shorttitle {MW Star Formation}
\shortauthors{Vutisalchavakul et al.}

\begin{document}

\setcounter{table}{0}

\title{Star Formation Relations in the Milky Way}
\author{
Nalin Vutisalchavakul\altaffilmark{1},
Neal J. Evans II\altaffilmark{1}, 
Mark Heyer\altaffilmark{2}
}
\altaffiltext{1}{Department of Astronomy, The University of Texas at Austin,
2515 Speedway, Stop C1400, Austin, Texas 78712-1205, U.S.A.}
\altaffiltext{2}{Department of Astronomy, University of Massachusetts, Amherst, Massachusetts 01003, U. S. A.}
\email{nje@astro.as.utexas.edu}

\begin{abstract}
The relations between star formation and properties of molecular clouds
are studied based on a sample of star forming regions in the Galactic Plane. 
\added{Sources were selected by having radio recombination lines to
provide identification of associated molecular clouds and dense clumps.
Radio continuum and mid-infrared emission were used to determine star 
formation rates, while \coo\ and submillimeter dust continuum emission
were used to obtain masses of molecular and dense gas, respectively.}
\replaced{This was accomplished mainly by two approaches:  exploring empirical relations between star formation rate
 and properties of molecular clouds in the data; and 
testing some previously proposed models or hypotheses about
star formation.}{We test whether total molecular gas or dense gas
provides the best predictor of star formation rate. We also test two
specific theoretical models, one relying on the molecular mass divided
by the free-fall time, the other using the free-fall time divided by
the crossing time. Neither is supported by the data.} 
The data are also compared to those from nearby
star forming regions and extragalactic data. The star formation
``efficiency," defined as star formation rate divided by mass, spreads
over a large range when the mass refers to molecular gas\added{; the
standard deviation of the log of the efficiency decreases by a factor
of three when the mass of relatively dense molecular gas is used
rather than the mass of all the molecular gas.}
\deleted{The spread is much reduced if the mass of relatively dense molecular
gas is used. 
Models that introduce free-fall time and/or dynamical time
are not particularly effective.}

\end{abstract}


\section{Introduction}

Given the importance of star
formation in the evolution of galaxies, understanding the regulation
of star formation is crucial. 
Early work on star formation on galaxy
scales relied on empirical star formation laws
\citep{Schmidt:1959,Kennicutt:1989}, 
with little connection
to the detailed studies of star formation in our own Galaxy.
Recently there has been more focus on integrating the understanding
of the process of star formation from the scale of galaxies to the
much smaller scales of regions within molecular clouds
\citep{Kennicutt:2012,Kruijssen:2014,Krumholz:2014}. 

While large scale
studies provide essential information on the relation between
large scale properties of galaxies and star formation, the process of
converting gas into stars takes place on a smaller scale. Since molecular clouds (MCs) 
are the sites of 
star formation in galaxies, it is essential to establish the key processes and 
sequences within molecular 
clouds that regulate the production of newborn stars in order to gain a deeper 
understanding of processes at galactic scales. 
While there are some recent high-spatial
resolution studies of nearby galaxies that can resolve regions of MCs 
(e.g., \citealt{2007ApJ...661..830R, 2013ApJ...779...46H}), 
the Milky Way offers the highest resolution view 
to investigate the connection
between star formation and the local gas properties. 

There are several recent surveys of dust and molecular line emission in the 
Milky Way that provide information on the distributions and 
properties of \gmc s. 
Ideally, star formation in \gmc s can be directly evaluated by identifying
stars or young-stellar objects inside the clouds, which along with the
information on their mass and lifetime provide a good estimate of star
formation rate (SFR) for the clouds. This direct method of estimating
SFR has been applied for nearby ($d < 830$ pc) molecular clouds
\citep{Heiderman:2010,Gutermuth:2011,Lada:2010,Evans:2014}, 
but these have a limited range in properties, making it difficult to test
theories for the importance of cloud properties in controlling star
formation rates. Furthermore, they are primarily low-mass 
($\mean{\mcloud} \sim 3000$ \msun) clouds 
\citep{Heiderman:2010}
whose star formation does not fully sample the IMF. 
Their star formation activity 
would be almost entirely invisible to observers in other galaxies.
The goal of this paper is to extend this effort to larger clouds
where massive stars are formed, both to sample a larger range of
cloud properties and to examine regions more comparable to those that
can be observed in other galaxies.

The challenge is that the more massive clouds with more fully sampled
IMFs are all quite distant; even Orion does not sufficiently sample the
IMF to use the extragalactic indicators of star formation rate
\citep{Kennicutt:2012}. 
For those distant clouds, counting YSOs is very difficult, both because
of sensitivity limits and because of background source confusion
\citep{Dunham:2011}, although recent work has been more successful
\citep{Heyer:2016}. 
To study star formation in a larger sample of Galactic \gmc s, 
we resort to indirect tracers of
SFR, such as those commonly used in extragalactic studies including
H$\alpha$, UV continuum, total infrared luminosity, mid-infrared
emission, and radio continuum emission. 
The shorter wavelength tracers (H$\alpha$, UV continuum) cannot be
used in the plane of the Galaxy because of dust obscuration, and
the total far-infrared luminosity awaits full release of surveys
with {\it Herschel}. In this study, we use mid-infrared and radio continuum
emission.
It is known that these indirect tracers derived from extragalactic
data are problematic when applied to smaller regions such as \gmc s.
The problem arises mostly from the assumptions of a fully-sampled 
IMF and a star formation history that is constant over a long
timescale
\citep{Kruijssen:2014,Krumholz:2015}.
Several recent studies  of SFR tracers
in regions with different properties suggest that some tracers
offer reasonable measures of SFR (although still with large scatter) in regions
above a certain minimum SFR
\citep{Wu:2005,Vutisalchavakul:2013}.

In this paper, we collect data from surveys of radio recombination lines,
radio continuum, and mid-infrared emission to measure SFR, 
\coo\ spectroscopy to evaluate MC properties, and  
millimeter dust continuum emission to trace the dense gas component. 
We describe these data
sets in \S \ref{data} and summarize our selection of star forming
regions and their association with gas in \S \ref{analysis}. 
Various models for star formation prediction are tested with these data 
in \S \ref{tests}.  In  \S \ref{lowcomp},  we
compare our results to similar studies of nearby clouds, and in
\S \ref{exgal}, we put our results into the context of studies of other
galaxies.


\section{Data}\label{data}

\subsection{Radio Recombination Lines and \hii\ Region Catalog}
\citet{Anderson:2014}, hereafter A14, compiled a catalog of \hii\ regions
within the Milky Way.
The A14 catalog comprises 
over 8000 Galactic \hii\ regions and \hii\ region candidates identified 
by mid-infrared (MIR) emission morphology using WISE 12 $\mu$m and 22 $\mu$m 
data. Many of these sources are associated with radio recombination lines
(hereafter RRLs)
H86$\alpha$ through H96$\alpha$ \citep{Anderson:2011}
or H$\alpha$ emission, confirming their association with \hii\ regions. 

\subsection{Radio Continuum Data}
Radio continuum emission closely associated with recent star formation 
in the Milky Way
comes from free-free emission of ionized gas around high-mass
 stars.
The Very Large Array Galactic Plane Survey (VGPS),
observed the \ion{H}{1}  spectral line and 21~cm 
radio continuum emission from the 
first Galactic quadrant covering the Galactic
longitude between $18^\circ < l < 67^\circ$ with varying Galactic
latitude range from $|b| < 1.3^\circ$ to $2.3^\circ$ 
at the resolution of 1\arcmin\ \citep{Stil:2006}.  To provide sensitivity on 
larger scales, short spacing data obtained with the Green Bank Telescope were
added to the interferometric continuum data.

\subsection{Mid-Infrared Data}
The Wide-field Infrared Survey Explorer (WISE) 
mapped the entire sky in four IR bands at
3.4, 4.6, 12, and 22 \mic\  \citep{Wright:2010}.  
The MIR band at 22 \mic, which has a resolution of 12\arcsec,
provides a measure of star formation rate. 
The WISE Image Atlas provides image tiles covering 1.564$^\circ \times $
x 1.564$^\circ$ in area with 1.375\arcsec\ per pixel.
Larger 22 \mic\ image mosaics 
covering the segment of the Galactic Plane in this study were generated 
from the set of image tiles 
using the MONTAGE mosaic software \citep{Jacob:2010}. The
characteristic saturation level for the point sources, defined to be the level 
at which 50\% of
the sources have some saturated pixels, in the 22 \mic\ band is
about $12$ Jy. 
For a few bright sources that are saturated in the WISE 22 \mic\ band, we
use the data from the Infrared Astronomical Satellite (IRAS)
25 \mic\ band. The IRAS Improved Reprocessing of the IRAS Survey
(IRIS) provides images of the sky at 25 \mic\ band at the resolution
of $\approx 4$\arcmin\ \citep{Miville:2005}. 
The resolution of IRAS 25 \mic\ data is considerably
lower than that of WISE 22 \mic\ band; therefore, we use IRAS 25
\mic\ images for sources saturated in WISE 22 \mic\ only if 
the source is resolved with no confusion with other nearby
sources. 

\subsection{\co\ \jj10\ Emission}
Data from the Boston University-FCRAO Galactic Ring Survey (GRS) and the Exeter-FCRAO 
Galactic Plane Survey (EXFC) provide data on
molecular clouds.
The GRS
surveyed the  \co\ \jj10\ emission between 
Galactic longitudes 
$18^\circ < l < 55.7^\circ$ and Galactic latitudes, $|b|<1^\circ$ 
using the Five College Radio Astronomy Observatory
14-m telescope \citep{Jackson:2006}.
The LSR
velocity (\vlsr) coverage is from $-5$ to 135 km s$^{-1}$ for $l \le 40^\circ$
and from $-5$ to 85 km s$^{-1}$ for $l > 40^\circ$. The final data cubes
are gridded at 22\arcsec\ and at velocity resolution of 0.21 km
s$^{-1}$.   The median main beam sensitivity of the GRS is 0.21~K.
The EXFC survey covered \cotw\ and \coo\ \jj10\ emission within two areas:
$55.7^\circ < l < 102.5^\circ$, 
$|b|< 1^\circ$ and $141^\circ < l < 192^\circ$, $-3.5^\circ < b < 5.5^\circ$ 
with the FCRAO 14~m telescope 
\citep{2016ApJ...818..144R}. 
The median main beam sensitivity at 110~GHz in the area of this study is 0.46~K.
For both the GRS and EXFC surveys, the half power beam width of the telescope at 110 GHz is 48\arcsec. 
All of the \co\ data used in this study (GRS and EXFC) have been post-processed to remove 
contributions from the antenna error beam. 

\subsection{Millimeter Dust Continuum Emission}
The dense gas within molecular clouds is examined using 
1.1 mm dust emission from the 
Bolocam Galactic Plane
Survey (BGPS). The BGPS observed part of the Galactic Plane 
at the effective resolution of 33\arcsec\
 and $1\sigma$ sensitivity ranging from 30 to 100 mJy per beam \citep{Aguirre:2011, Ginsburg:2013}. 
 The BGPS provides contiguous
 observational data over the ranges of $|b| < 0.5^\circ$ for 
$-10.5^\circ < l < 90.5^\circ$, $|b| < 1.5^\circ$ for $75.5^\circ < l < 87.5^\circ$, 
and additional coverages over selected regions
described in \citet{Aguirre:2011} and \citet{Ginsburg:2013}.

Owing to the removal of atmospheric signal by spatial filtering, 
the data from BGPS recovers most of the astrophysical emission out to
the scale of 80\arcsec\ and partially recovers emission out to the scale of
300\arcsec\ \citep{Ginsburg:2013}.  The spatial filtering subtracts signal from 
molecular clouds that subtend solid angles greater than 300\arcsec. 
For these reasons, the data are sensitive to compact regions within clouds with 
enhanced 
column densities, and likely, volume densities.  

The BGPS version 2.1 data release includes a source catalog extracted from 1.1 mm
maps using a seeded watershed algorithm \citep{Rosolowsky:2010, Ginsburg:2013}.  
The Bolocat V2.1 consists of 
8594 sources with 1.1 mm flux from photometry
(background-subtracted) for the aperture radii of 20\arcsec, 
40\arcsec, and 60\arcsec, as well as flux integrated over the source area.  

All the Bolocat sources in version 1 of the BGPS catalog within the Galactic
longitude range of $7.5^\circ < l < 194^\circ$ were followed up with 
spectroscopic observations of dense gas tracers, using the
 HCO$^+$ and N$_2$H$^+$ 3-2 
transitions with the Arizona Radio Observatory Submillimeter 
Telescope \citep{Schlingman:2011,Shirley:2013} 
with 51\% of the sources detected in at least 
one of the molecular lines. Additional molecular line observations
of NH$_3$ (1,1), (2,2), and (3,3)  inversion lines are available for 
sources in the inner Galaxy from the Green Bank Telescope
\citep{Dunham:2011}.
The molecular line observations of the Bolocat
sources provide kinematic information to link the dust continuum source to its 
parent molecular cloud and the \hii\ regions.


\section{ANALYSIS}\label{analysis}

In this section we describe the selection criteria for the sample of 
star forming regions (\S 3.1), the calculations of star formation rates
(\S 3.2), and the extraction and measurement of the properties
of MCs and dense gas (\S 3.3).

\subsection{The Samples}
Studies of star formation in the plane of the Galaxy are challenging
because it is non-trivial to clearly associate a star forming event, identified by broad-band 
photometry,  with
a particular molecular cloud. 
In this study, the detection of a hydrogen radio recombination line 
(RRL) arising
within an \hii\ region is required as a marker of star formation
because we need velocity information to tie the star formation to
the molecular gas.  
Target sources 
with well defined distances 
are selected from the compilation of \hii\ regions with a RRL by A14.
First, we restrict our source list to Galactic longitudes, $30^\circ < l < 63^\circ$ to 
sample a volume of the disk that is external to the stellar bar but crosses the 
Scutum and Sagitarius spiral arms and extends to the Vulpecula Rift.   
To ensure coverage by the GRS, whose minimum velocity is $-5$ \kms, 
the LSR velocities of the radio recombination lines 
(\vrrl) are restricted to be greater than 0~\kms. 
 A consequence of 
excluding negative velocities in this longitude range is the exclusion of star forming 
regions on the far side of the Galaxy with Galactic radii greater than 8.5~kpc. 
The primary sample \added{of 66 sources} 
comprises the RRLs that could be associated with \gmc s. 
The positions, velocities, and distances of the selected sample are 
listed in Table~\ref{table1}. \added{The sample size is reduced to 51
sources after a minimum star formation rate is imposed (\S 3.2.3).}

When testing relations between dense  gas and star 
formation properties, we further restricted the sample to
sources that can be linked to one or more sources from the 
BGPS survey of 1.1 mm dust continuum emission sources 
\citep{Ginsburg:2013} 
and follow-up spectral line observations in \hcop\ and \nthp\
\citep{Schlingman:2011,Shirley:2013}.  
To augment the sample, we include a subset of  BGPS sources 
studied by \citet{Battisti:2014} that are also associated with 
RRL \hii\ regions. 
\added{The resulting sample size for dense gas relations is 44.}  

\begin{deluxetable*}{lccccccc}
\tablecaption{Radio Recombination Line Sources
\label{table1}}
\tabletypesize{\scriptsize}
\tablewidth{0pt}
\tablehead{ 
\colhead{WISE Name} & \colhead{$l$} & \colhead{$b$} & \colhead{\vrrl} & 
\colhead{$D$}  & \colhead{$\sigma(D)$} & \colhead{$R_{\rm GAL}$} & 
\colhead{Ref.}\tablenotemark{a} \\
\colhead{} & \multicolumn{2}{c}{(deg)} & \colhead{(\kms)} & \multicolumn{2}{c}{(kpc)} & \colhead{(kpc)} & \colhead{}
}
\startdata
G031.561+00.376 &   31.561 &    0.377 &  24.3 &  13.0 &  0.4 &  6.9 & 2,2 \\
G032.077-00.230 &   32.077 &   -0.229 &  97.7 &   8.2 &  1.4 &  4.7 & 2,2 \\
G032.030+00.048 &   32.030 &    0.049 &  91.1 &   7.2 &  1.2 &  4.8 & 1,1 \\
G032.272-00.226 &   32.272 &   -0.226 &  21.5 &  12.8 &  0.7 &  7.1 & 1,1 \\
G032.473+00.204 &   32.473 &    0.204 &  47.0 &  11.2 &  1.7 &  6.1 & 1,1 \\
G032.582+00.001 &   32.590 &   -0.001 &  77.4 &   9.4 &  1.8 &  5.2 & 1,1 \\
G032.587-00.330 &   32.587 &   -0.330 & 103.8 &   7.2 &  1.3 &  4.6 & 1,1 \\
G032.733+00.209 &   32.733 &    0.209 &  16.1 &  13.2 &  1.4 &  7.3 & 1,1 \\
G032.870-00.427 &   32.870 &   -0.427 &  50.6 &  10.9 &  1.7 &  6.0 & 1,1 \\
G033.419-00.005 &   33.419 &   -0.004 &  76.5 &   9.3 &  1.8 &  5.3 & 1,1 \\
G033.643-00.229 &   33.643 &   -0.228 & 102.9 &   7.1 &  1.1 &  4.7 & 1,1 \\
G033.809-00.190 &   33.809 &   -0.190 &  40.3 &  11.4 &  0.9 &  6.4 & 1,1 \\
G033.941-00.039 &   33.942 &   -0.039 &  57.2 &  10.4 &  1.1 &  5.8 & 1,1 \\
G034.041+00.052 &   34.041 &    0.053 &  36.4 &  11.6 &  0.7 &  6.5 & 1,1 \\
G034.089+00.438 &   34.089 &    0.438 &  32.6 &  11.8 &  1.3 &  6.7 & 1,1 \\
\enddata
\vspace{0.4cm} 
\tablenotetext{a}{First and second number cites the reference for the 
\vrrl\ and distance values respectively. 
1:\citep{Anderson:2014}, 2:\citep{Battisti:2014}, 3:\citep{Ellsworth:2015b} }
\tablecomments{Table 1 is
published in its entirety in
the electronic edition of
the Astrophysical Journal.
A portion is shown here
for guidance regarding its
form and content.}
\end{deluxetable*}

The choice of requiring \hii\ regions means that all the star-forming
regions associated with molecular clouds in this study are forming high-mass
stars, making the use of infrared and radio continuum tracers less
problematic. 
The sample is clearly biased against starless
clouds or even clouds with low-level star formation similar to that in
solar-neighborhood clouds.

\subsubsection{Assigning Distances }
Distances to most of the target RRL regions are derived by A14 using 
trigonometric parallax of associated masers or more typically, kinematic distances 
 assuming the rotation curve of the Milky Way derived by \citet{Brand:1993} 
and resolving the distance ambiguity for the inner Galaxy.  
For 9 sources where distances for the RRL \hii\ regions are not defined by A14, 
the cloud distance is assigned to the mean of the
distances from the associated Bolocat sources, where distances were
obtained from \citet{Ellsworth:2015b}, who adopted the 
rotation curve of \citet{Reid:2009}. \added{For the 9 sources taken from \citet{Battisti:2014}, 
we adopt the distance used in their study that assumed the rotation curve of \citet{Clemens:1985}. }
The difference in distances using the \replaced{two}{three} 
rotation curves is small compared to other uncertainties with the kinematic distance method. 
\added{The mean fractional distance uncertainties, $\sigma(D)/D$, for objects in our sample 
 are 0.15 for A14, 0.19 for \citet{Ellsworth:2015b}, and 0.20 for \citet{Battisti:2014}. }
The
molecular clouds are assigned distances of the \hii\ regions with which these are associated.
For sources without
a spectroscopic dense gas tracer, velocities were obtained by
connecting molecular gas observations from the GRS \co\ data
to Bolocat sources \citep{Ellsworth:2015b}. The combination of
dense gas tracers and \co\ data resulted in 45\% of the Bolocat
V2.1 sources with assigned \vlsr.  
The distance uncertainties are 
estimated from the combined uncertainties in the choice of Galactic
rotation curve, the streaming motions of 7 km s$^{-1}$, and the solar
circular rotation speed \citep{Anderson:2012, Anderson:2014}.

The sources span a large range of heliocentric distances of
1.7 to 13.2~kpc, with an average distance of $8\pm3$ kpc and a median
of 8.1 kpc. The span of  Galactocentric radii ($R_{\rm GAL}$) is less, ranging
from 4.6 to 8~kpc with an average of $6.3\pm 0.8$~kpc and a median of 6.1~kpc.
They should be more representative of star formation 
activity in the Galaxy than nearby, well-studied targets such as those
in the Gould Belt
\citep{2015ApJS..220...11D}. 

\subsection{Star Formation Rates}

The radio continuum at 21 cm and MIR emission at 22 or 25 \mic\ are the star
formation tracers used in this study.
We refer to them as ``radio" and ``MIR" in what follows.
Using continuum emission as a tracer creates a problem of source 
confusion between different emitting regions along the line of
sight. If more than one emitting region lies within the
angular vicinity of a MC, one cannot distinguish which regions are associated 
with the MC. 
To mitigate the problem of confusion along the line of
sight, the RRL data are used to 
associate radio continuum emission and MIR emission with molecular gas
\added{
by requiring velocity agreement with the molecular gas (see \S 3.2.1).
}

\floattable
\begin{deluxetable*}{lcccc}
\tablecaption{Star Formation Rates
\label{table2}}
\tabletypesize{\scriptsize}
\tablewidth{0pt}
\tablehead{ 
\colhead{WISE Name} &  \colhead{$\sfr_{\rm radio}$} & \colhead{$\sigma(\sfr_{\rm radio})$} & \colhead{$\sfr_{\rm MIR}$} & \colhead{$\sigma(\sfr_{\rm MIR})$}  \\
\colhead{}    & \multicolumn{2}{c}{(M$_\odot$/Myr)}    &   \multicolumn{2}{c}{(M$_\odot$/Myr)}
}
\startdata
G031.561+00.376 & 120.30 & 12.03 & 31.01 &  3.10 \\
G032.077-00.230 & 58.60 &  5.86 & 150.80 & 15.08 \\
G032.030+00.048 & 45.18 &  4.52 & 119.79 & 11.98 \\
G032.272-00.226 & 39.06 &  3.91 & 34.12 &  3.41 \\
G032.473+00.204 &  9.94 &  0.99 & 24.05 &  2.41 \\
G032.582+00.001 & 11.64 &  1.16 &  6.98 &  0.70 \\
G032.587-00.330 & 12.60 &  1.26 & 10.71 &  1.07 \\
G032.733+00.209 & 148.94 & 14.89 & 176.60 & 17.66 \\
G032.870-00.427 & 35.57 &  3.56 & 49.42 &  4.94 \\
G033.419-00.005 & 45.24 &  4.52 & 63.96 &  6.40 \\
G033.643-00.229 &  6.86 &  0.69 &  0.00 &  0.00 \\
G033.809-00.190 & 30.38 &  3.04 & 82.71 &  8.27 \\
G033.941-00.039 & 89.57 &  8.96 & 41.53 &  4.15 \\
G034.041+00.052 & 28.04 &  2.80 & 206.83 & 20.68 \\
G034.089+00.438 & 24.02 &  2.40 & 40.01 &  4.00 \\
\enddata
\vspace{0.4cm} 
\tablecomments{Table 2 is
published in its entirety in
the electronic edition of
the Astrophysical Journal.
A portion is shown here
for guidance regarding its
form and content.}
\end{deluxetable*}

\subsubsection{Star Formation Rates Derived from Radio Continuum Emission}
For each of the RRL locations,
a region of associated radio continuum
 emission is defined using the following steps.
First, the radio emission is fitted to a 
2D-Gaussian profile, centered on the RRL position. If the emission is reasonably well fitted 
with the 2D-Gaussian, the source is labeled as a compact
source with sizes ($\sigma_x, \sigma_y)$ 
of the fitted Gaussian.  The radio flux is calculated from aperture photometry 
with an aperture radius $3(\sigma_x+\sigma_y)/2$ and
an appropriate sky annulus.  If the radio 
emission could not be fitted well with a 2D Gaussian (which is the
case if the source is extended or in a crowded region), then a 
polygon is used  to define the region of the source emission and another
region for estimating background emission.
The RRL sources with no significant radio continuum  emission above the
background are excluded from further analysis.

If more than one RRL source is found within the solid angle of the radio
aperture and 
RRL velocities are within $\pm 5$ \kms\ of each other, then the RRL 
sources are considered as a single  star-forming region. 
If the velocities differ by more than 5 \kms\ and the 
radio emission cannot be separated, then the sources are excluded 
from our target list. 

Radio emission from galaxies is comprised of both synchotron and 
free-free emission 
components with synchotron emission dominating at 21~cm \citep{Condon:1992}.
In contrast, the higher spatial resolution in the Milky Way surveys favors
localized regions of free-free emission.
Since our targets are selected by the detection of the hydrogen 
recombination line, we assume the origin of the 21~cm signal is 
free-free emission from an \hii\ region 
excited by ionizing far-UV radiation from massive stars. 
Therefore, we do not use the extragalactic relation between radio emission
and star formation, which is primarily a relation between synchrotron
emission on large scales and star formation averaged over about 100 Myr
\citep{Kennicutt:2012}.
Instead, we use the compact free-free radiation from the \hii\ region, which,
like H$\alpha$, averages over about 3 Myr 
\citep{2011ApJ...741..124H,Kennicutt:2012}.

The free-free
emission is related to the ionizing luminosity by the expression 
\begin{eqnarray}
 \frac{N_{\rm UV}}{\text{phot s}^{-1}}  = 6.3 \times 10^{52} 
 \left(\frac{T_e}{10^4 \text{K}}\right)^{-0.45}  
 \left(\frac{\nu}{\text{GHz}}\right)^{0.1} 
   \nonumber \\
  \times  
 \left(\frac{L_{\rm T}}{10^{20}\text{W Hz}^{-1}}\right), 
\end{eqnarray}
where $N_{\rm UV}$ is the number of Lyman continuum photons per second,
$T_e$ is the electron temperature, $\nu$ is the frequency, and $L_{\rm T}$
is the thermal emission luminosity assuming optically thin gas \citep{Condon:1992}. 
For an electron temperature of $10^4$ K, and an IMF described by 
\added{\citep{Chomiuk:2011}}, the SFR is calculated from
\begin{equation}
\text{SFR(radio)} 
= 0.47 \times 10^{-14}
\left( \frac{\nu}{\text{GHz}} \right)^{0.1} \times 
\left ( \frac{L_{\rm T}}{\text{W Hz}^{-1}} \right) \;\;{\msun \text{Myr}^{-1}} 
\end{equation}
\citep{Chomiuk:2011,Vutisalchavakul:2013}.
\explain{Reference corrected}

This expression may underestimate the SFR if UV photons are not
absorbed, as in the case of a density bounded \hii\ region.
\added{If a Salpeter slope is continued down to 0.1 \msun, the
derived SFRs would be 1.44 times higher \citep{Chomiuk:2011}.}  
The star formation rates derived from radio emission for our 
sample of RRL regions are listed in Table~\ref{table2}.

\subsubsection{Star Formation Rates Derived from MIR Emission}
The process used to define the
MIR-emitting region associated with a RRL source is similar to the
method used for radio continuum images described in the previous section. 
The MIR emission at the location of the  radio continuum region is 
examined. If the distribution of MIR emission is compact and isolated,
it is fitted
 to a 2D Gaussian profile.  Otherwise we define the emitting
 region with a polygon. Photometry for all of the MIR emitting
 regions is performed similarly to the radio continuum regions. 
Most of the sources that are compact in radio images are also compact in 
MIR emission. 
In the case of G033.643-00.229, the background subtracted flux is less than zero.
This target is excluded from subsequent analyses.

The SFR is calculated from the extragalactic
relation \citep{Calzetti:2007}: 
\begin{equation} \label{eq:ssfr}
\sfrmir  
= 1.27{\times}10^{-32} \times
[L_{24} (\text{ergs s}^{-1})]^{0.885} \;\;(M_\odot \ \text{Myr}^{-1})
\end{equation}
where $L_{24}$ is the 24\mic\ luminosity. 
\added{The IMF assumed by \citet{Calzetti:2007} is consistent with
that used for the SFR from the radio continuum emission.}  
 Here, we substituted the WISE 22 \mic\ or the IRAS 25 \micron\
luminosity for the 
24 \mic\ luminosity.
The corresponding star formation rates derived from the mid-infrared luminosity 
are listed in Table~\ref{table2}. 

\begin{figure*}[h]
\includegraphics[scale=0.5]{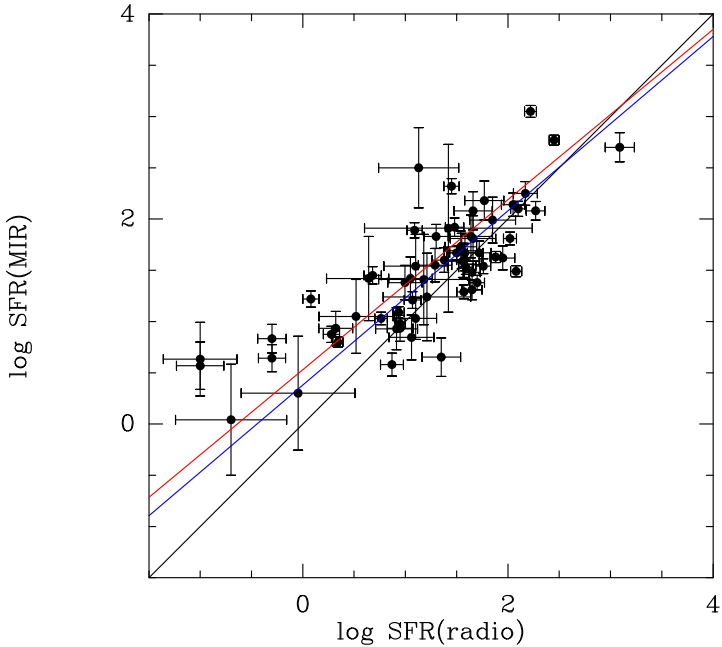}
\caption{The comparison between SFR(MIR) and SFR(radio)\added{, measured
in \msunmyr}. The solid
black line represents the 1-1 line where the two SFRs are equal. The
blue line is a fit to the data, including uncertainties in both axes.
The red line is fit to a smaller sample of sources from
\citet{Vutisalchavakul:2013}.
}
\label{fig:compare_sfr}
\end{figure*}

\subsubsection{Comparing SFR from radio continuum and MIR}

The two different estimations of SFR are compared in 
Figure~\ref{fig:compare_sfr}. Because the luminosities of both tracers
are proportional to a flux times the square of the distance, the uncertainties 
are computed by propagating a nominal 10\% uncertainty in the flux measurement
and the distance uncertainty in Table~\ref{table1} (see the Appendix for a
description of error propagation). 
The distance uncertainty is usually, but not always, dominant.
The SFRs from the two tracers are
comparable, with the SFR from MIR usually being somewhat
higher, especially at low SFRs. The black solid line in 
Figure~\ref{fig:compare_sfr} is a line of equality, while the blue solid
line is a fit to the data, using $\chi^2$ minimization with errors in both
variables. The fit is
\begin{equation}
\log[{\rm SFR(MIR)}] = a + b\ \log[{\rm SFR(radio)}]
\end{equation}
with $a = 0.38\pm0.04$ and $b = 0.85\pm0.02$. The slope is nearly
identical to that found by analysis of a more limited sample,
while the coefficient is a bit less, but within the uncertainties:
($a = 0.53\pm0.17$ and $b = 0.83\pm0.08$)
\citep{Vutisalchavakul:2013}; 
that fit is shown as a red line in Figure~\ref{fig:compare_sfr}.

The slightly sub-linear fits and the larger discrepancies at low
SFR can result from the fact that the radio emission is more
sensitive to the upper end of the IMF (as is also true for H$\alpha$
or any diagnostic requiring ionized gas), as can be seen from Table 1 of 
\citet{Kennicutt:2012}.
For this reason, we use SFR(MIR) in further analysis. This tracer also
begins to underestimate the SFR for SFRs less than about 5 \msunmyr, 
corresponding to a total far-infrared luminosity of $10^{4.5}$ \lsun\
(\citealt{Vutisalchavakul:2013, Wu:2005}). Consequently,
we limit further investigation to those sources with SFR above this
value\added{, leaving 51 sources}.

Even above the threshold of 5 \msunmyr, SFRs may be poorly estimated. The
dominant source of uncertainty in SFR is  the conversion from the
star-formation tracers to SFR. The common method used to calculate the
conversion factors is to use a stellar population synthesis
model, stellar evolutionary models, and atmospheric models to connect
stellar populations to their photometric output. Then assumptions of
a fully-sampled IMF and a star-formation history (SFH)
connect the SFR of a region to the light output for each tracer
\citep{Leitherer:1999}. 
The effect of the stochastic sampling of the IMF and SFH on
the reliability of the SFR tracers has been a topic of many recent
studies \citep{Fumagalli:2011, daSilva:2012,daSilva:2014, Kruijssen:2014,Krumholz:2015}. 
\citet{daSilva:2014} used the 
Stochastically Lighting Up Galaxies (SLUG) simulations to study the
effect of IMF and SFH sampling on the conversion from photometric
observations of the star-formation tracers to SFR. Instead of a unique
SFR for a given luminosity (in the tracers), the effect of IMF and SFH
sampling results in probability distributions of SFR, which
can have a large bias and scatter for regions with low SFR. 
The result from SLUG, given an input of a constant SFR on the timescale
of 500 Myr, for the bolometric luminosity as a SFR tracer gives a
scatter in the log(SFR) of $\approx 0.6$ dex in the SFR range of $10 -
100$ \msunmyr, assuming uncertainties in the flux of 0.25
dex \citep{daSilva:2014}.  
Consistent with that analysis,
the SFR(MIR) appears to underestimate SFR for the Milky Way
 by a factor of $\approx 2-3$ \citep{Chomiuk:2011}.  
Because the effect is usually to underestimate the SFR,
we treat this as a systematic uncertainty,
rather than including it in the individual SFRs.

\subsection{Gas Structures}

\subsubsection{Molecular Cloud Properties}

The linking of a molecular cloud to the radio recombination line \hii\ regions is similar to the method 
employed by \citet{Battisti:2014} to connect larger clouds to 1.1 mm continuum sources.
A 100$\times$100~pc$^2\times$40 km s$^{-1}$ data cube centered on the $l$,
$b$, \vrrl\ 
 coordinates of the \hii\ region is extracted from the larger CO survey.  For a handful of nearby sources, the 100~pc 
edge may lie outside the Galactic latitude boundary of the survey, in which case the area is truncated to the largest available 
square area. 
 The segmentation program, CPROPS \citep{Rosolowsky:2006}, 
is applied to these extracted data to identify sets (islands) of contiguous voxels 
with CO brightness temperatures above a given 
threshold value.  This threshold is varied depending on the complexity of the local background emission generated by unrelated 
foreground and background molecular 
clouds with comparable radial velocities with the intent to determine the lowest threshold value that distinguishes a
structure from this background emission.  We do not examine substructure within these islands of 
emission. 

\begin{figure*}[h]
\includegraphics[scale=0.75]{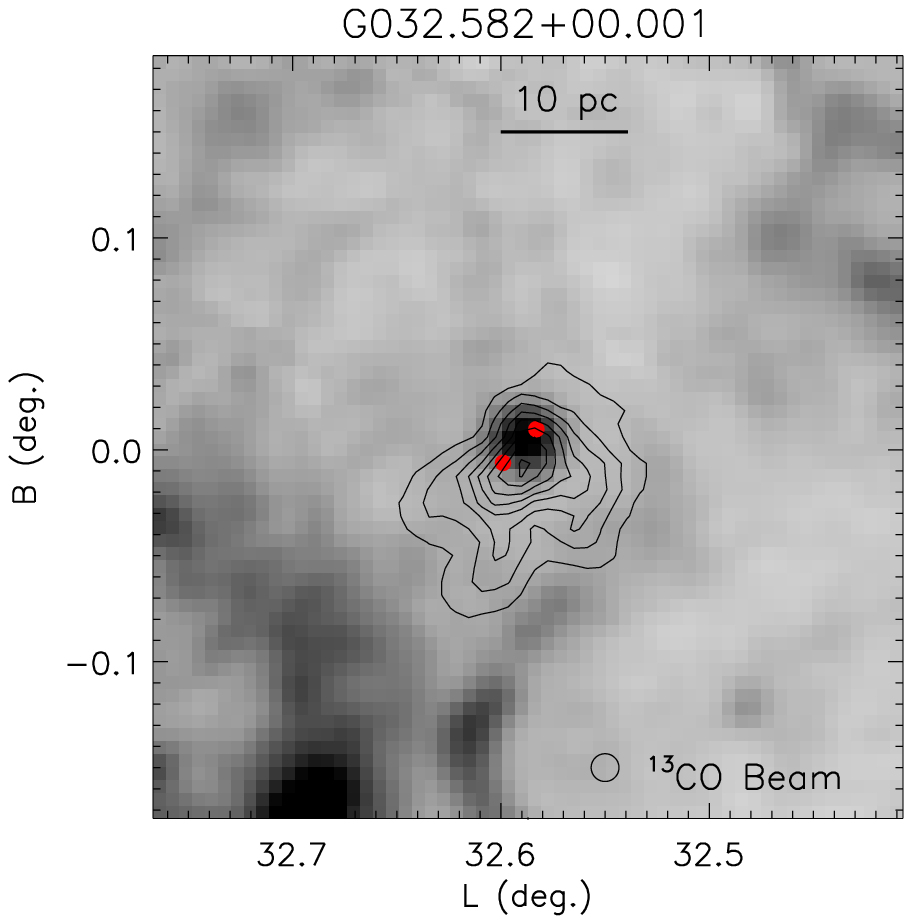}
\includegraphics[scale=0.75]{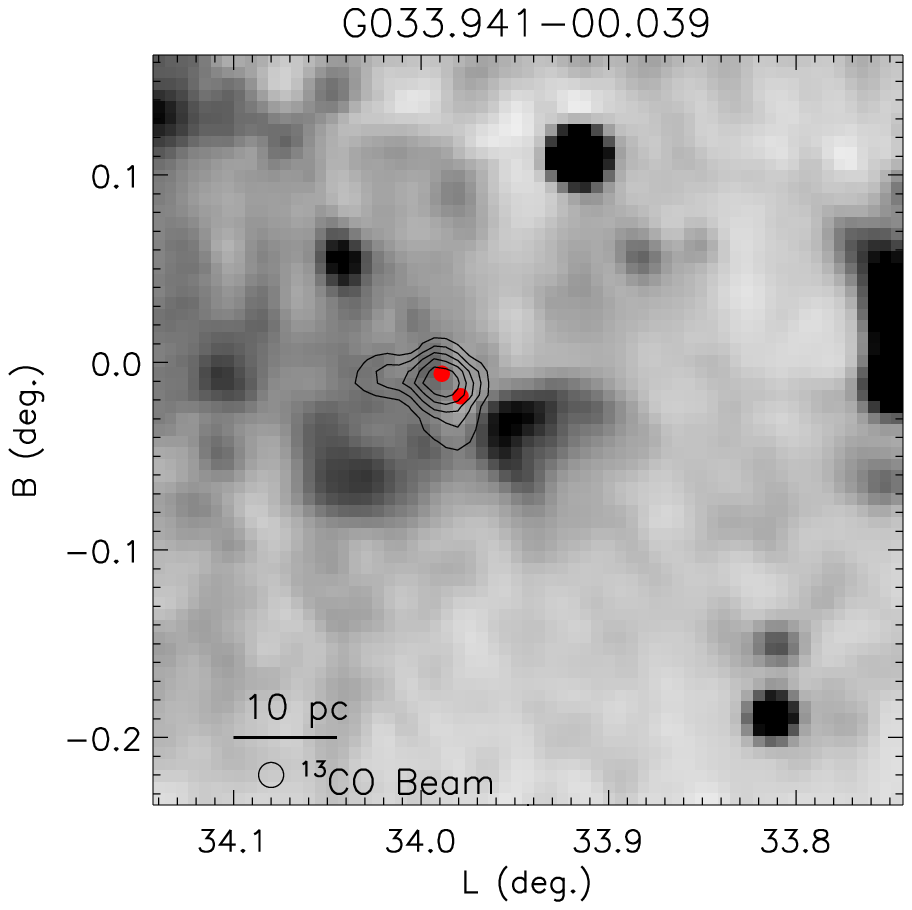}
\caption{Images of integrated \co\ \jj10\ emission (contours) overlayed on 
21~cm continuum emission for RRL targets
(left) G032.582+00.001 and (right) G033.941-00.039.  
Contours range from 1 to 15 K \kms\ spaced by 2 K \kms\ 
for
G032.582+00.001 and 1 to 5 K \kms\ spaced by 1 K \kms\ for G033.941-00.039. 
The radio continuum source to the southwest of the cloud is the associated source.
The solid, red circles mark the locations of associated Bolocam sources of dust 
continuum emission. 
\deleted{These are examples of Group~1 (left) and Group~2 (right) clouds.}
}
\label{co_clouds}
\end{figure*}

For a given threshold, hundreds of structures may be identified by the segmentation algorithm distributed 
over the 10$^4$ pc$^2$ area and 40~\kms\ range.  To narrow the search for the molecular cloud associated with the \hii\ region, we 
select a subset of this list with the condition, 
$|v_{\rm CO}-v_{\rm RRL}| < 10$ \kms, where $v_{\rm CO}$ is the 
velocity centroid of the CO structure. 
For each structure in this subset, the distribution of velocity-integrated CO emission and its spatial 
relationship to the \hii\ region are examined.
A CO structure is assigned to the \hii\ region if its boundaries enclose the \hii\ region or 
\replaced{its distribution is strongly 
peaked towards the \hii\ region position.}
{if there is a local intensity maximum coincident with 
the \hii\ region position.}
\replaced{If these conditions are not satisfied,  a structure may also be linked if 
its CO boundary, 
is contiguous to the \hii\ region}
{A structure may also be linked if the solid angle of the detected 
radio continuum emission from the 
HII region partially overlaps with the solid angle of the CO emitting region or if the 
respective boundaries of the radio continuum and CO emissions are within 2\arcmin\ (2 times the FWHM beam width of the radio continuum data) at some 
point along their perimeters. 
}
In several cases,  no structure can be confidently assigned to the \hii\ 
region position, so we select 
\replaced{a nearby structure that could be reasonably associated with the \hii\ region. }
{the structure with the most comparable velocity to the RRL velocity 
that is within 5 \coo\ FWHM beam widths of the HII region position. }
Two examples of the cloud extraction are shown in Figure~\ref{co_clouds}.

Once a cloud is linked to the \hii\ region by the above criteria, its properties are determined from the set of voxels
that comprise the structure.  These properties include intensity-weighted positional moments and 
\coo\ luminosity.  
While the voxels that comprise the cloud are defined by the applied threshold, the cloud properties such as mass, size, 
and velocity dispersion, are 
extrapolated from this brightness temperature threshold to a hypothetical cloud edge at a brightness temperature of 0~K.  
The properties and uncertainties of each MC linked to an RRL source are summarized in Table~\ref{table3}.

\floattable
\begin{deluxetable*}{lccccccccccccccc}
\tabletypesize{\scriptsize}
\rotate
\tablecaption{$^{13}$CO Derived Molecular Properties
\label{table3}}
\tablewidth{0pt}
\tablehead{ 
\colhead{WISE Name} & \colhead{Type} & \colhead{$r_{cloud}$} & \colhead{$\sigma(r_{cloud})$} & \colhead{${\delta}v$} & \colhead{$\sigma({\delta}v)$} &
\colhead{\mcloud}  & \colhead{$\sigma(\mcloud)$} & \colhead{\mvir} & \colhead{$\sigma(\mvir)$} & 
\colhead{$\alpha$} & \colhead{$\sigma(\alpha)$} &
\colhead{\ncloud} & \colhead{$\sigma(\ncloud)$} & \colhead{$\tff$} & \colhead{$\sigma(\tff)$} \\
\colhead{}    & \colhead{}  & \multicolumn{2}{c}{(pc)}  &  \multicolumn{2}{c}{(km s$^{-1}$)} & \multicolumn{2}{c}{(10$^3$ M$_\odot$)} &
\multicolumn{2}{c}{(10$^3$ M$_\odot$)} &
\colhead{} & \colhead{} & \multicolumn{2}{c}{(cm$^{-3}$)} & \multicolumn{2}{c}{(Myr)} 
} 
\startdata
G031.561+00.376 &  2 & 15.81 & 8.57 & 1.30 & 0.23 &  66.3 &  26.8 &  31.1 &  20.1 & 0.47 & 0.36 &  60 & 105 &  4.1 &  7.1 \\
G032.077-00.230 &  3 & 17.67 & 2.88 & 2.63 & 0.77 & 105.0 &  42.4 & 142.1 &  86.3 & 1.35 & 0.99 &  69 &  53 &  3.9 &  3.0 \\
G032.030+00.048 &  2 & 27.21 & 0.01 & 1.88 & 0.01 & 180.6 &  72.9 & 111.8 &   1.2 & 0.62 & 0.25 &  32 &  19 &  5.6 &  3.3 \\
G032.272-00.226 &  1 & 9.37 & 0.07 & 1.41 & 0.08 &   7.8 &   3.2 &  21.7 &   2.5 & 2.76 & 1.16 &  34 &  20 &  5.5 &  3.2 \\
G032.473+00.204 &  1 & 15.56 & 0.05 & 2.06 & 0.03 & 101.7 &  41.1 &  76.8 &   2.2 & 0.75 & 0.31 &  97 &  57 &  3.2 &  1.9 \\
G032.582+00.001 &  1 & 10.41 & 0.06 & 1.84 & 0.10 &  40.4 &  16.3 &  41.0 &   4.5 & 1.01 & 0.42 & 129 &  76 &  2.8 &  1.7 \\
G032.587-00.330 &  1 & 10.06 & 0.02 & 2.28 & 0.03 &  20.7 &   8.3 &  60.8 &   1.6 & 2.94 & 1.19 &  73 &  43 &  3.7 &  2.2 \\
G032.733+00.209 &  1 & 12.47 & 0.02 & 2.66 & 0.02 &  93.3 &  37.6 & 102.6 &   1.6 & 1.10 & 0.44 & 173 & 102 &  2.4 &  1.4 \\
G032.870-00.427 &  1 & 13.56 & 0.02 & 1.59 & 0.03 &  72.5 &  29.3 &  39.8 &   1.5 & 0.55 & 0.22 & 105 &  62 &  3.1 &  1.8 \\
G033.419-00.005 &  1 & 15.40 & 0.04 & 2.10 & 0.04 &  48.8 &  19.7 &  78.9 &   3.0 & 1.62 & 0.66 &  48 &  28 &  4.6 &  2.7 \\
G033.643-00.229 &  3 & 24.88 & 0.01 & 2.85 & 0.01 & 232.3 &  93.8 & 234.9 &   1.7 & 1.01 & 0.41 &  54 &  32 &  4.3 &  2.6 \\
G033.809-00.190 &  1 & 5.02 & 0.11 & 1.68 & 0.12 &  23.1 &   9.3 &  16.5 &   2.4 & 0.71 & 0.31 & 657 & 390 &  1.2 &  0.7 \\
G033.941-00.039 &  2 & 8.63 & 0.23 & 1.75 & 0.23 &  21.4 &   8.7 &  30.7 &   8.1 & 1.43 & 0.69 & 120 &  72 &  2.9 &  1.7 \\
G034.041+00.052 &  1 & 13.94 & 0.02 & 1.56 & 0.03 &  33.0 &  13.3 &  39.4 &   1.5 & 1.20 & 0.48 &  44 &  26 &  4.8 &  2.8 \\
G034.089+00.438 &  1 & 7.39 & 0.04 & 1.47 & 0.04 &  12.2 &   4.9 &  18.6 &   1.0 & 1.52 & 0.62 & 109 &  64 &  3.1 &  1.8 \\
\enddata
\vspace{0.4cm} 
\tablecomments{Table 3 is
published in its entirety in
the electronic edition of
the Astrophysical Journal.
A portion is shown here
for guidance regarding its
form and content.}
\end{deluxetable*}

For \co\ data, column densities are derived following the expressions in \citet{Pineda:2010}. Here, we assume 
optically thin emission, so column density in the upper ($J=1$) rotational energy level is 
\begin{equation}
N_{1,{\rm thin}}=
\frac{8{\pi}k_B\nu_{10}^2}{hc^3A_{10}C(\tex)}\int T_{\rm B}(v)dv
\end{equation}
where $k_B$ is the Boltzmann constant, $h$ is the Planck constant, 
$c$ is the speed of light, and
$\nu_{10}$, $A_{10}$, $T_{B}$, \tex\ 
 are the frequency, Einstein A coefficient,
brightness temperature, and excitation temperature
 for the \coo\ \jj10\ transition. 
The factor, $C(\tex)$, contains temperature dependent terms, 
\begin{equation}
C(\tex)=\left(1-\frac{exp(T_{\rm L}/\tex)-1}{exp(T_L/T_{\rm bg})-1}\right)
\end{equation}
where $T_{\rm L}=h\nu_{10}/k_B$ and $T_{\rm bg}$=2.725~K.  
The total column density of \coo\ is 
\begin{equation}
N(\coo)=N_{1,thin}Z exp(h{B_\circ}J(J+1)/{k_B}\tex)/(2J+1)
\end{equation}
where $B_\circ$ is the rotational constant for \coo, and $Z$ is the partition function, which accounts for population in other states
\begin{equation}
Z = \sum_{J=0}^\infty (2J+1) exp(-h{B_\circ}J(J+1)/k_B\tex) 
\end{equation}
Finally, to convert to molecular hydrogen column density, 
$N(\hh)=N(\coo)[\hh/\coo]$,
we assume a constant excitation temperature of 8~K and a H$_2$ 
to $^{13}$CO abundance ratio of 4.1$\times$10$^{5}$ for all clouds.
This value for \tex\ is based on studies by \citet{Heyer:2009} and \citet{Roman-Duval:2010} who 
found mean \tex\ values of 7-8~K derived from \cotw\ data for large samples of clouds.  
The assigned abundance value is based on mean abundance values for 
nearby clouds using infrared-derived extinction as a proxy for 
\hh\ column density \citep{Lada:1994,Pineda:2008, Pineda:2010, Ripple:2013}. 
This leads to a conversion between \co\ integrated intensity and molecular hydrogen column density of 
3.9$\times$10$^{20}$ cm$^{-2}$/(K km s$^{-1}$) that is a required argument in the CPROPS 
program to calculate mass.  The cloud mass is then 
\begin{equation}
\mcloud = 3.9{\times}10^{20} \mu m_H D^2 \int d\Omega \int dv T_B(l,b,v)
\end{equation}
where $D$ is the distance and the integrals are over the solid angle, $\Omega_{cl}$, and velocity range for the 
set of cloud voxels. 

There are several sources of 
uncertainty in the MC mass.
These include
distance, 
abundance and $T_{ex}$ variations, and random errors owing to 
the observations.  This last component is small compared to the other sources and is ignored. 
For \co, we 
estimate the uncertainty of the conversion factor from integrated intensity 
to $N(\coo)$
by varying the assumed excitation temperature from 4 to 16~K in steps of 2~K. The fractional root mean square of 
values 
about the adopted value for 8~K is 27\%.  The variations of \co\ abundance, 
$[\hh/\coo]$ both within clouds and from cloud-to-cloud in the 
solar neighborhood is $\sim$30\% \citep{Ripple:2013}. 
These uncertainties are included in the values in Table \ref{table3},
but distance errors are not (see Appendix).

Assuming a spherical cloud and a power law density profile with index of 1, 
the CPROPS package calculates the virial mass, \mvir, from the expression, 
\begin{equation}
\mvir = 1040{\delta}v^2 \rcloud ~M_\odot 
\end{equation}
where ${\delta}v$ is the 1-dimensional velocity dispersion in \kms\ and \rcloud\ is the effective 
radius of the cloud in parsecs. 
\added{The coefficient can be found in
\citet{1988ApJ...333..821M} after converting from FWHM to $\delta v$.} 
The virial parameter, $\alpha =\mvir/\mcloud$, offers a coarse measure 
of the boundedness of the cloud, assuming there is no significant external pressure. 
The mean volume density of the cloud, $\ncloud$, is derived from the 
\mcloud\ and \rcloud, assuming a spherical cloud geometry, 
$\ncloud = 3\mcloud/4{\pi}{\mu}m_H \rcloud^3$.

\floattable
\begin{deluxetable*}{lcccccchhhh}
\tabletypesize{\scriptsize}
\tablecaption{BGPS-Derived Dense Gas Properties
\label{table4}}
\tablewidth{0pt}
\tablehead{ 
\colhead{WISE Name} & \colhead{\mdense} & \colhead{$\sigma(\mdense)$} & 
\colhead{\rdense} & \colhead{$\sigma(\rdense)$}  & \colhead{DGMF} & 
\colhead{$\sigma(\rm DGMF)$} & 
\nocolhead{$n_D$} & \nocolhead{$\sigma(n_D)$} & \nocolhead{$\tau_{ff,D}$} & 
\nocolhead{$\sigma(\tau_{ff,D})$} \\
\colhead{} & \multicolumn{2}{c}{(\msun)} & \multicolumn{2}{c}{(pc)} & 
\multicolumn{2}{c}{} & 
\nocolhead{(cm$^{-3}$)} &
\nocolhead{(cm$^{-3}$)} &
\nocolhead{(Myr)} &
\nocolhead{(Myr)} 
}
\startdata
G031.561+00.376 & 1143.2 & 663.3 &  0.8 &  0.1 & 0.017 & 0.012 & 7174.4 & 5268.7 &  0.4 &  0.3 \\
G032.077-00.230 & 2503.7 & 1337.0 &  4.0 &  0.6 & 0.024 & 0.016 & 146.3 & 102.1 &  2.6 &  1.8 \\
G032.030+00.048 & 15865.8 & 6177.0 &  5.6 &  0.8 & 0.088 & 0.049 & 323.3 & 192.4 &  1.8 &  1.1 \\
G032.272-00.226 & 3099.3 & 1254.4 &  3.4 &  0.5 & 0.396 & 0.226 & 296.0 & 179.2 &  1.9 &  1.1 \\
G032.473+00.204 & 3942.9 & 1548.8 &  5.5 &  0.8 & 0.039 & 0.022 & 87.1 & 52.0 &  3.4 &  2.0 \\
G032.582+00.001 & 2351.8 & 1179.0 &  4.4 &  0.7 & 0.058 & 0.037 & 97.9 & 66.0 &  3.2 &  2.2 \\
G032.587-00.330 & 150.8 & 118.8 &  2.3 &  0.3 & 0.007 & 0.006 & 43.2 & 39.2 &  4.9 &  4.4 \\
G032.733+00.209 & 23918.2 & 9357.0 &  6.5 &  1.0 & 0.256 & 0.144 & 320.9 & 191.4 &  1.8 &  1.1 \\
G032.870-00.427 & 5439.9 & 1554.3 &  5.8 &  0.6 & 0.075 & 0.037 & 100.4 & 43.1 &  3.2 &  1.4 \\
G033.419-00.005 & 5440.5 & 2134.5 &  5.3 &  0.8 & 0.112 & 0.063 & 133.9 & 80.0 &  2.8 &  1.7 \\
G033.643-00.229 & -99.9 & -99.9 & -99.9 & -99.9 & -99.900 & -99.900 & -99.9 & -99.9 & -99.9 & -99.9 \\
G033.809-00.190 & 7411.8 & 2225.8 &  6.9 &  0.7 & 0.321 & 0.162 & 82.8 & 36.3 &  3.5 &  1.5 \\
G033.941-00.039 & 2954.6 & 1155.2 &  4.2 &  0.6 & 0.138 & 0.077 & 140.8 & 83.9 &  2.7 &  1.6 \\
G034.041+00.052 & 2547.2 & 1033.1 &  6.5 &  1.0 & 0.077 & 0.044 & 32.8 & 19.9 &  5.6 &  3.4 \\
G034.089+00.438 & 4360.4 & 1741.7 &  4.3 &  0.7 & 0.357 & 0.203 & 192.1 & 115.6 &  2.3 &  1.4 \\
\enddata
\vspace{0.4cm} 
\tablecomments{This table is
published in its entirety in
the electronic edition of
the Astrophysical Journal.
A portion is shown here
for guidance regarding its
form and content.}
\end{deluxetable*}

\subsubsection{Dense Gas Mass} 
Dense gas properties are estimated for each MC using source fluxes and sizes from 
the BGPS Version 2.1 source catalog. 
The relation between gas mass and dust continuum flux is 
given by the expression
\begin{equation}
\md = \frac{S_{1.1} D^2 (\rho_g/\rho_d)}{B_\nu (T_{\rm dust}) 
\kappa_{\rm dust, 1.1}}, 
\end{equation}
where  $S_{1.1}$ is the 1.1 mm flux density, $D$ is the distance,
 $\kappa_{\text{dust},1.1}$ 
is the dust opacity at 1.1 mm per dust mass,  assumed to be 
 1.14 cm$^2$ g$^{-1}$\citep{Ossenkopf:1994}, 
and $\rho_g/\rho_d$ is the gas-to-dust mass ratio, taken to be
100 \citep{Hildebrand:1983}. 
Using this expression, we derive the mass and associated uncertainties using a Monte Carlo 
simulation that assumes  Gaussian errors on the flux density and a 
Gaussian distribution of dust temperatures with a mean of 20~K and 
a standard deviation of 8~K \citep{Battersby:2011, Ellsworth:2015b}.
The total dense gas mass for each MC is estimated from the sum 
of the masses for all Bolocam sources
within the cloud mask with and without \vlsr\ information. 
We also derive  the dense gas mass fraction (DGMF) corresponding to the ratio of the mass 
traced by BGPS 1.1 mm sources to the mass of the
MC as traced by \co\ emission, 
\begin{equation}
DGMF = \frac{\md}{\mcloud}. \nonumber
\end{equation}
Properties of the dense gas component are given in Table \ref{table4}.


\section{Classification of MCs}\label{classification}

We classified our sample of star forming  MCs into three distinct groups
based on the locations of the associated \hii\ regions with respect to the 
molecular cloud. 
Group 1 MCs have associated radio continuum emission whose solid angle is
 mostly circumscribed by the boundary of \co\ emission.
There are 37 MCs in Group 1.
For Group 2 MCs, 
the radio continuum emission partially 
overlaps or is contiguous with \co\ boundaries but the RRL sources lie 
outside the clouds. 
\added{The left panel of figure \ref{co_clouds} shows a Group 1 cloud, while the right
panel shows a Group 2 cloud.}
There are 21 MCs in Group 2. 
Finally, 
Group 3 clouds are MCs in which the radio continuum emission is well displaced 
from the \co\ boundaries. 
The association of the MC with the \hii\ region is most uncertain for
Group 3 sources. 
With no 
angular overlap with the MC, the association is solely based on
the chosen spatial and velocity offsets. There are only 8 MCs in this
group. Due to the small number and the uncertainty of associating
gas and star formation in this group, we
exclude Group 3 MCs from all further analysis. 

The classification into groups in this study has some
overlap with the classification used by \citet{Fukui:1999}
and \citet{Kawamura:2009} (K09),
but our spatial resolution is higher than that of K09 so we examine 
associations between MCs and star formation on smaller
spatial scales but also require similar velocities to account for line of sight confusion. 
Nevertheless, our classifications likely correspond to a similar 
evolutionary
sequence in which Group 1 is the earliest stage of massive star formation
(Type II of K09), Group 2 describes the initial feedback from 
massive stars, and Group 3 is a very late stage in 
which the original cloud has been mostly photoionized or dispersed, leaving only small fragments, similar to Type III clouds in the K09 classification.

To explore whether Group 1 and Group 2 sources differ in the main
properties we are considering, we computed averages and standard
deviations (in the log for quantities with large ranges) for the two
groups; these are shown in Table \ref{grouptab2}. The two groups have very 
similar averages for \mcloud, \mdense, \ncloud, \rcloud, and
SFE (SFR/\mcloud).
Group 2 clouds may have somewhat larger mean values for $\alpha$ and somewhat smaller values for DGMF. Because the differences are not statistically 
significant, we do not distinguish the two groups in the rest of the analysis.

\floattable
\begin{deluxetable*}{lcccc}
\tablecaption{Comparison of Groups 1 and 2 
\label{grouptab2}}
\tablewidth{0pt}
\tablehead{
\colhead{Property} &
\colhead{Group 1 Average} & 
\colhead{Group 2 Average} & 
\colhead{Group 1 Number} &
\colhead{Group 2 Number}  
 }
\startdata
\mean{\log \mcloud} & $4.58\pm0.48$  & $4.43\pm0.57$  &  $32$ & $19$ \\
\mean{\log \mdense} & $3.49\pm0.98$  & $3.19\pm1.55$  &  $30$ & $14$ \\
\mean{\log \ncloud} & $1.97\pm0.38$  & $2.01\pm0.63$  &  $32$ & $19$ \\
\mean{\rcloud} & $12.6\pm6.6$  & $12.5\pm8.2$  &  $32$ & $19$ \\
\mean{\alpha} & $1.44\pm1.11$  & $1.86\pm1.26$  &  $32$ & $19$ \\
\mean{DGMF} & $0.12\pm0.11$  & $0.06\pm0.045$  &  $30$ & $14$ \\
\mean{\log SFE} & $-2.93\pm0.54$  & $-2.89\pm0.54$  & $32$ & $19$ \\
\enddata
\end{deluxetable*}

\section{Testing Star Formation Relations}\label{tests}

The question we address in this section is the following:
which properties of the ISM best predict the SFR for the clouds
in this sample?
The first proposed relation \citep{Schmidt:1959,Schmidt:1963} 
suggested that the SFR would be a function of the density of gas.
For extragalactic studies, the surface densities of 
star formation rate (\ssfr) and gas ($\Sigma_{gas}$) are
easier to measure than are volume densities.
Studies of other galaxies found that $\ssfr \propto \Sigma_{gas}^{1.4}$ 
provided a good fit for a wide range of galaxies \citep{Kennicutt:1998}. 
When only molecular gas is considered, linear relations are generally
found ($\ssfr \propto \Sigma_{gas}$)
(\citealt{2008AJ....136.2846B,2013AJ....146...19L}),
even for the outer regions of galaxies, where the gas is predominantly
atomic 
\citep{2011AJ....142...37S}.
Studies of galaxies on sub-kiloparsec scales have shown that
star formation is associated with molecular gas
(e.g., \citealt{2013AJ....146...19L}). 
Within the Milky Way, studies with sub-parsec resolution show
that star formation is further concentrated within relatively
dense or opaque parts of molecular clouds
(\citealt{Lada:2010,Heiderman:2010}).
 
In this section, we use our sample of star forming regions
to test some of the proposed models for how SFRs are determined in
\gmc s.  In doing so, we consider integrated properties, such
as cloud mass, mass of dense gas, or cloud mass per free fall time 
as variables in the following equation, where $X$ represents the variable in 
question.
\begin{equation}
\sfr = A X^n
\end{equation}
Because of the large range of parameters, and to be consistent with
previous studies, we generally fit the logarithmic version of these relations.
\begin{equation}
\log \sfr = \log A +n \log X
\label{eq:sfl_sk2}
\end{equation}

We examine the various relations
that are used in extragalactic work or that have been proposed by
theorists for clouds in the Milky Way selected to have substantial
star formation.
 We use the SFR
from the mid-infrared data and consider only clouds with a SFR
of at least 5 \msunmyr, as discussed earlier.
The basic sample thus includes 51 sources. The samples for other
variables are somewhat smaller depending on which data are available.
For the dense gas relations, only 44 sources are available because not
all clouds had BGPS sources.
We calculate the Pearson correlation coefficient to evaluate the probability 
that 
the measured relationship could emerge from a population of random numbers. 
If the correlation is significant \added{(using Pearson $r$, a $3 \sigma$
correlation requires $|r| > 3/\sqrt{N_s - 1}$, where $N_s$ is the sample
size)}, we fit the data to the model 
in equation \ref{eq:sfl_sk2} to determine the parameters, 
$\log(A)$ and $n$, considering errors in both variables using the MPFITEXY routine  
\citep{Williams:2010}, which 
depends on the MPFIT package \citep{Markwardt:2009}.
In this model-fitting program, the intrinsic scatter of the data about the model is 
included in the error-weighting of the data.  If necessary, this intrinsic 
scatter is iteratively modified so that the reduced $\chi^2$ value of the fit is about unity.
The scatter of points is characterized by the root mean square of the displacement 
of log(SFR) values from the best fit line for each value of X. 
The parameters from the fits are given in Table \ref{tab:fits}.
The column labeled DOF shows the sample size minus the number (2) of
fit parameters.


\begin{deluxetable*}{lcccccc}
\tablecaption{Star Formation Rate Fits
\label{tab:fits}}
\tablewidth{0pt}
\tablehead{
\colhead{Variable} &
\colhead{$\log A$} & 
\colhead{$n$} & 
\colhead{DOF} &
\colhead{$\chi^2$}  &
\colhead{RMS}  &
\colhead{Pearson $r$}  
 }
\startdata
Cloud Mass & $-1.38\pm0.56$  & $0.66\pm0.12$  & 49 & 0.997 & 0.46 & $0.49$ \\
Dense Mass & $-2.12\pm0.62$  & $1.09\pm0.18$  & 42 & 1.019 & 0.52 & $0.58$ \\
\mcloud/\tff & $-5.61\pm1.60$  & $1.80\pm0.39$  & 48 & 0.919 & 0.75 & $0.55$ \\
\tff/\tdyn\tablenotemark{a}  & \nodata    & \nodata  & \nodata & \nodata & \nodata & $0.13$ \\
\enddata
\tablenotetext{a}{In this case the quantity being fitted is \epsff, rather
than \sfr }
\end{deluxetable*}

\subsection{Star Formation Rate versus Cloud Mass } 

\begin{figure*}[h]
\center
\includegraphics[scale=0.4]{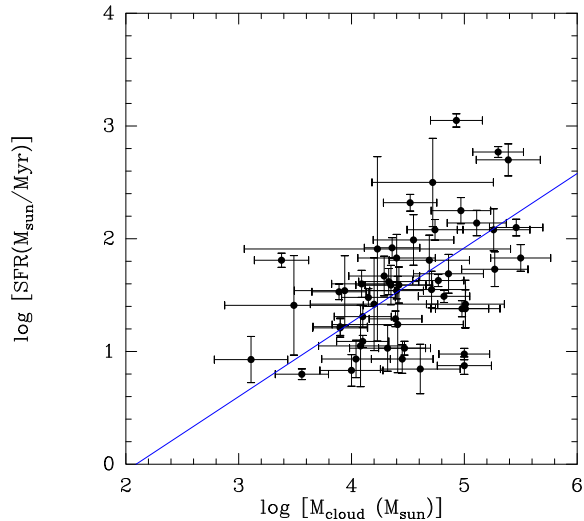}
\caption{The relation between SFR and \mcloud. 
The  line is the linear least-square fit to the data. }
\label{fig:sfl_sk}
\end{figure*}

The first and the simplest proposition is that SFR depends solely on
the total mass of the MC (\mcloud). 
This is the extragalactic star formation relation
applied on the cloud scale.
In this case $X = \mcloud$, as determined by the \coo\ emission.
Figure~\ref{fig:sfl_sk} shows the SFR versus \mcloud.
The dotted lines represent the linear least-square fits to the log of the data.
The fit parameters to Equation~\ref{eq:sfl_sk2} gives the slope 
\added{$n = 0.66\pm0.12$, sub-linear by nearly 3 $\sigma$}
(see Table \ref{tab:fits}).
The relation for the Galactic Plane clouds is thus considerably
flatter than the linear relation found for other galaxies, 
\added{averaging over kpc scales} 
(\citealt{2008AJ....136.2846B,2013AJ....146...19L,2011AJ....142...37S}).

\subsection{Star Formation Rate versus Dense Gas Mass}

A tighter relation has been observed in other galaxies
between SFR and dense gas mass. \citet{Gao:2004} 
observed a tight linear relationship between $L(\rm IR)$ and
$L$(HCN)  in a sample of infrared galaxies. \citet{Wu:2005} 
found that the relationship is extended to the scale of
Galactic dense clumps. These studies and the study of nearby 
molecular clouds led to \replaced{a proposed model}{the
proposal} that molecular gas above some  density can 
form stars efficiently \citep{Goldsmith:2008, Lada:2010, Heiderman:2010}. 
The star-formation threshold model \citep{Lada:2012, Evans:2014}
states that the SFR in \gmc s is \replaced{determined}{best predicted}
 by the amount of dense gas above some
 threshold density.   In this case, $X = \mdense$, as measured
by the sum of masses of the BGPS sources within the cloud,
with the caveat that BGPS sources are lower density and more like
clouds at large distances owing to spatial filtering to remove atmospheric 
emission 
\citep{Dunham:2011}.

\begin{figure*}[h]
\center
\includegraphics[scale=0.4]{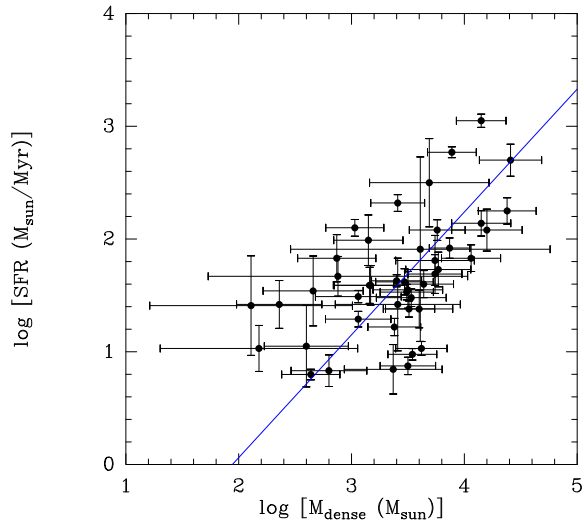}
\caption{The relation between SFR and \md. 
The line is the linear least-square fit to
the data. }
\label{fig:sfl_bgps}
\end{figure*}

Comparing SFR and \md\ in our sample yields a nearly linear relation 
($n = 1.09\pm 0.18$) with a Pearson correlation coefficient of 0.58.
Figure~\ref{fig:sfl_bgps} shows the result of the relation
between SFR and \md. 

\explain{Removed subsection on DGMF}

\subsection{SFR versus \mcloud/\tff }

The free-fall model states that the SFR depends on
 the molecular mass
over the free-fall time \citep{Krumholz:2005,Krumholz:2012}: 
\begin{equation}
SFR = f_{H_2} \, \epsff \times \frac{M_{gas}}{\tff}, 
\end{equation}
where $f_{H_2}$ is the fraction of molecular gas compared to the total gas
mass, $\epsff = \tff/\tdep$ is the star formation efficiency per
free-fall time, the gas depletion time is \tdep,
and the free-fall time is
\begin{equation}
\tff = \sqrt{\frac{3\pi}{32G\rho}}. 
\end{equation}
For \gmc s, where only molecular gas is concerned, the relation becomes
\begin{equation}
SFR = \epsff \times \frac{\mcloud}{\tff}.
\label{eq:sflwtff}
\end{equation}
Comparing data from nearby clouds to other galaxies, \citet{Krumholz:2012} estimated
an approximately constant $\epsff$ of 0.01.   This value is aligned with theoretical 
predictions \citep{Krumholz:2005,2012ApJ...759L..27P}. 

\begin{figure*}[h]
\center
\includegraphics[scale=0.5]{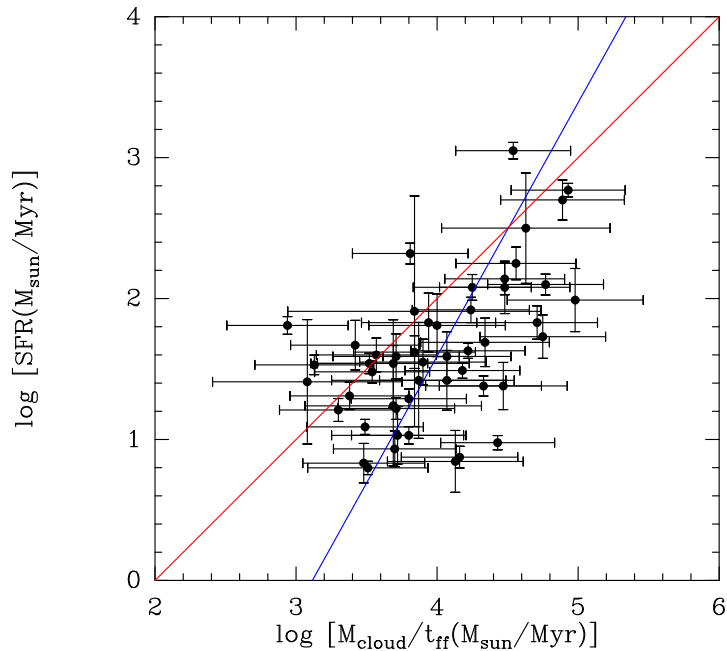}
\caption{The relation between SFR and mass over free-fall time.
The blue line is the linear least-square fit to
the data,
and the red line represents the free-fall model
with $\epsff = 0.01$.
}
\label{fig:sfl_tff}
\end{figure*}

We tested the free-fall
model with our cloud
sample by using the total \gmc\  mass from \co\ and using the
average density for \tff. 
 The result is shown in Figure~\ref{fig:sfl_tff}. 
The red line
represents the relation in Equation~\ref{eq:sflwtff} with
$\epsff = 0.01$. 
The fit to the data shows a \added{super-linear relation with $n = 1.80\pm0.39$.}
\deleted{very similar to that found for \mcloud.}
The Pearson correlation coefficient is 0.55. 
Generally speaking, including the free-fall time did not improve
the relation compared to using \mcloud, and the \replaced{sublinear}{super-linear} fit is
not consistent with the hypothesis of 
\citet{Krumholz:2012}.

\subsection{\epsff\  versus \tff/\tdyn}

As summarized by 
\citet{2014prpl.conf...77P},
simulations of turbulent molecular clouds indicate that the
speed of star formation, measured by \epsff, should depend on 
the virial parameter. In particular
 \citet{2012ApJ...759L..27P} show that turbulent simulations
closely follow the following relation.
\begin{equation}
\epsff = \epsilon_{\rm wind} e^{-1.6 \tff/\tdyn}
\end{equation}
where $\epsilon_{\rm wind}$ is the core to star efficiency, taken
to be about 0.5 due to winds removing material, and 
$\tdyn = \rcloud/\sigma_{\rm v, 3D}$, and 
$\sigma_{\rm v, 3D} = \sqrt{3} \sigma_{\rm v, 1D}$. The variable
$\tff/\tdyn$ is simply related to the virial parameter by
\begin{equation}
\tff/\tdyn = 0.86 \sqrt{\alpha}
\end{equation}
where $\alpha$ is the virial parameter.

\begin{figure*}[h]
\center
\includegraphics[scale=0.5]{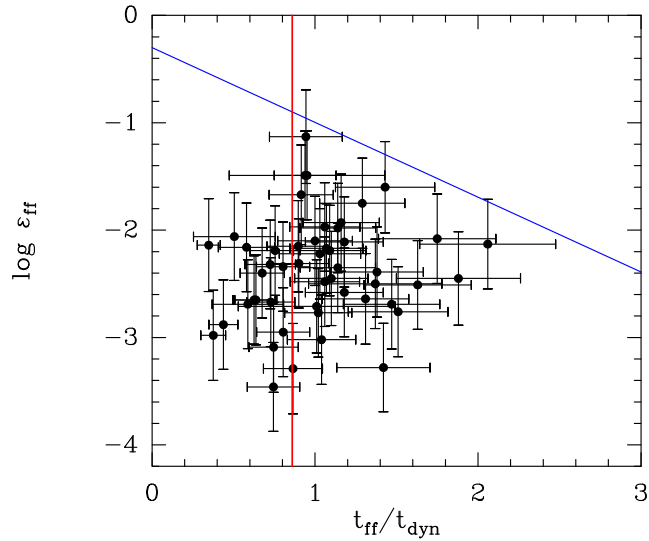}
\caption{The relation between \epsff\ and
free-fall time over dynamical (crossing) time.
The blue line is the prediction of the model of
\citet{2012ApJ...759L..27P}.
The red vertical line indicates $\alpha = 1$.
}
\label{fig:padoanmod}
\end{figure*}

The data are plotted in Figure \ref{fig:padoanmod},
along with the prediction of \citet{2012ApJ...759L..27P}.
The data show no significant correlation (Pearson correlation 
coefficient of $0.13$)
and lie on average a factor of 55 below the model
predictions.
\added{The mean efficiency \epsff\ for \mcloud\ is 0.008, comparable
 to the \citet{Krumholz:2012} value of 0.01. For this sample,
$\mean{\alpha} = 1.6\pm 1.2$; if we use that value in equations 17 and 18,
they would predict $\epsff = 0.09$, more than 10 times the observed
mean value.
}

\subsection{Summary of Star Formation Relationships}

Comparison of the values in Table~\ref{tab:fits} shows that all
three variables (\mcloud, \md, \mcloud/\tff) 
have similar correlation coefficients. 
\deleted{Using the rule of thumb that a $3 \sigma$ correlation requires
$r = 3/\sqrt{N-1}$, where $N$ is the number in the sample,} All the
correlations are significant at \replaced{that}{the $3 \sigma$} level.  Equivalently, we can 
exclude the null hypthosis that the correlations arise from a random 
distribution of variables  with a high level of probability.
Only the correlation with \mdense\ is consistent with a linear
relation. The fit to SFR versus \mcloud\ shows a slope much flatter
($0.66\pm0.12$)
than the usual values of $1.0$ to $1.5$.
Theory predicts a linear relation between SFE and \mcloud/\tff,
but the slope we find ($1.80\pm 0.39$) differs by more than
$2 \sigma$  from theoretical expectation.

Another theoretical prediction is that the star formation
efficiency per free-fall time should decrease exponentially with
the ratio of the free-fall time to the dynamical time. 
As shown in Figure~\ref{fig:sfl_tff}, 
there is no signficant correlation in our data and the values
lie well below the predictions.

Comparing SFR and \mcloud\, we found considerable scatter. 
This scatter has been observed and quantitatively explained in
several studies \citep{Onodera:2010, Schruba:2010, Kruijssen:2014}. 
One explanation is that the strong
correlation observed when looking at the scale of galaxies is due to
averaging over variations in different star forming regions'
properties. When looking at the scale of \gmc s, the properties of \gmc\ 
such as the evolutionary stages contribute to the scatter in the
relation between mass and SFR. Even in our sample where only sources
with overlap between molecular gas and star formation are selected
(thus leaving out the earlier stages such as IRDC and later stages
where stellar feedback disrupts the clouds), the scatter in SFR
and \mcloud\ relations is large. In previous work, we have shown that
averaging over larger regions in our Galaxy reduces the scatter
\citep{Vutisalchavakul:2014}.

\section{Comparison to Nearby Clouds}\label{lowcomp}

Now that we have extended the study of star formation relations to
more distant regions forming more massive stars in the Milky Way, we can compare
the results to those for nearby clouds
\citep{Evans:2014,Heiderman:2010}.
\added{
Those authors considered star formation relations {\bf within}
the nearby clouds, but we lack the resolution to do that for the
Galactic Plane clouds. In addition,
\citet{2013ApJ...778..133L}
highlighted the differences between star formation relations {\bf within}
and {\bf between} clouds, and
\citet{2016arXiv160507623P} has summarized the issues in comparing
surface densities of SFR and gas between nearby and more distant
clouds. 
Furthermore, plotting SFR versus mass introduces a correlation 
because both are proportional to the square of distance.
Cognizant of these issues, 
we focus on SFE (distance cancels out) 
and deal with the masses contained within
the entire cloud or a dense clump within the cloud, rather than
using surface density {\it within} a cloud or clump.
}

First, we consider whether comparable regions have been
selected. For the nearby clouds, a region above a threshold extinction,
generally $\av = 2$ mag, was selected for surveys of YSOs 
\citep{2009ApJS..181..321E}. 
For this paper, the clouds were defined by maps of \coo\ \jj10\ emission, followed by
extrapolation to zero emission. Our estimates of an effective threshold
for the Galactic Plane clouds range from $\av = 1$ to 3 mag.
This may be in addition to extinction of several magnitudes associated with layers
of atomic and molecular gas exterior to the \coo\ boundary limited by photodissociation. 
The mean density of the selected clouds for the Galactic Plane
sample is about 300 \cmv, lower than the average for the nearby
clouds of about 800 \cmv. 

The measure of ``dense" gas in the nearby
clouds was an extinction threshold of $\av = 8$ mag. Using the
typical rms noise in the BGPS survey of 0.1 mJy/beam 
\citep{Ginsburg:2013}, 
a dust temperature of 20 K, and OH5 dust opacities 
\citep{Ossenkopf:1994},
a 3 $\sigma$ detection limit translates to $\av = 7$ mag,
comparable to that used to define dense regions in the nearby clouds.
A second analysis, using the actual distribution of rms noise
in the BGPS yields a slightly smaller threshold, $\av = 4.6$ mag
at 90\% completeness (B. Svoboda, pers. comm.).

However, the mean volume density of the BGPS sources in the 
Galactic Plane sample is
only \eten3 \cmv, substantially lower than the mean density of
the regions above $\av = 8$ for the nearby clouds ($\mean{n} = 6\ee3$
\cmv). Thus, the galactic plane ``dense" gas is not as dense
as that defined in the nearby clouds.  For more distant targets, where the 
molecular clouds may subtend angles less than 3\arcmin, the mass and densities 
may reflect the full cloud rather than high density fragments \citep{Dunham:2011, Battisti:2014}.

A common feature of all observational studies of star formation is
a low star formation efficiency. Only a small fraction of
the gas in galaxies is actively forming stars. The fractional mass of
young stars in molecular clouds is around 5\%
(\citealt{2009ApJS..181..321E,2015ApJS..220...11D}).
In the extragalactic context, star formation efficiency (SFE) refers in
fact to the slow {\it pace} of star formation, the star formation
rate per unit mass, as reflected in \tdep, the inverse of SFE, being
1-2 Gyr. 
Parallel to our equations for SFR, we consider relations for the SFE
and \tdep\
\begin{equation}
\sfe = \sfr/X = A X^{n-1}
\end{equation}
\begin{equation}
\tdep = X/\sfr = A^{-1} X^{1-n}
\end{equation}
and their logarithmic versions.

\begin{figure*}
\center
\includegraphics[scale=0.6, angle=0]{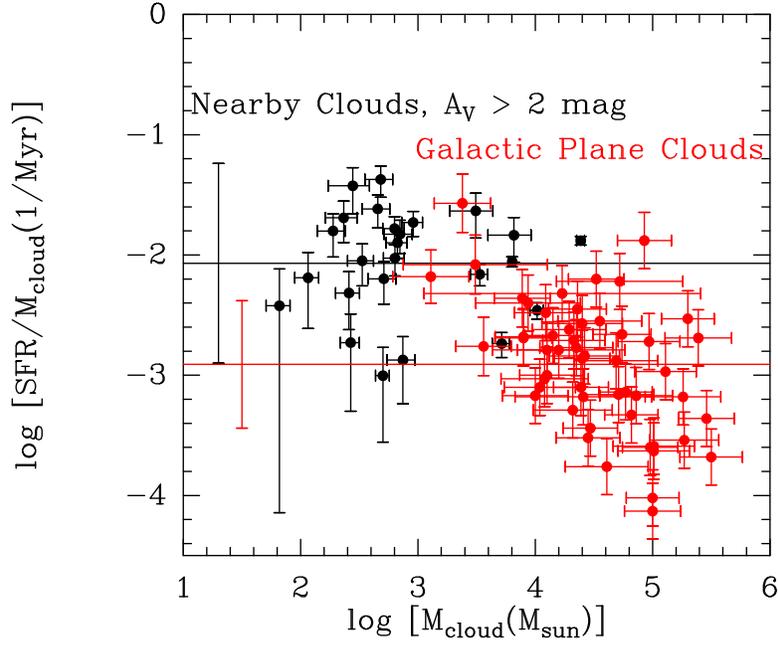}
\caption{
Plot of the logarithm of SFR per mass of molecular gas vs. the logarithm of
the mass of molecular gas. The red points are from this paper and the black
points are for nearby clouds, taken from 
\citet{Evans:2014}.
The black horizontal line is the mean value for the nearby clouds and
the red horizontal line is the mean value for the data in this paper.
The error bars at the far left are the standard deviations of $\log(\sfe)$.
}
\label{sfevsmass}
\end{figure*}

\begin{figure*}
\center
\includegraphics[scale=0.6, angle=0]{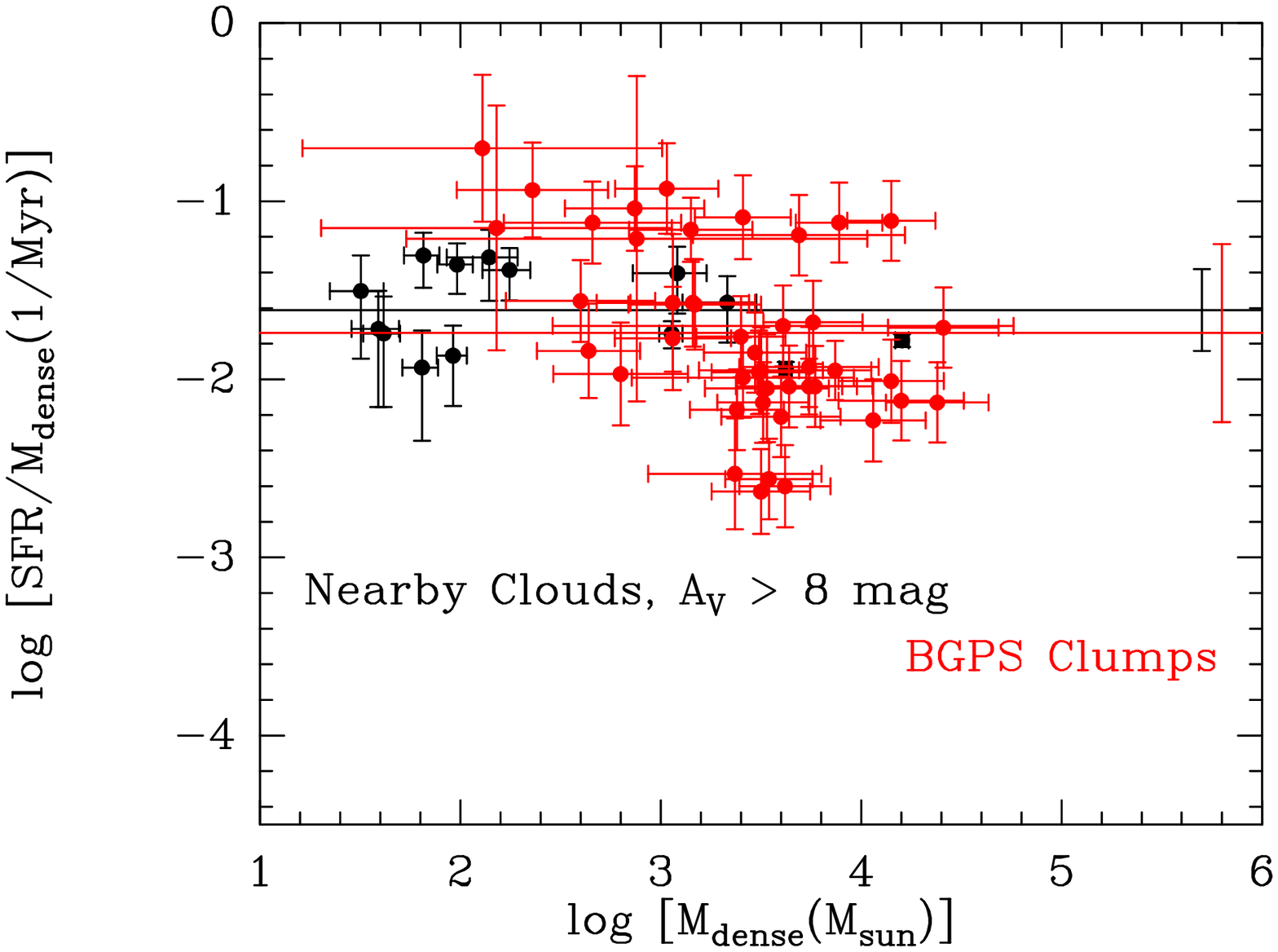}
\caption{
Plot of the logarithm of SFR per mass of dense gas vs. the logarithm of
the mass of dense gas. The red points are from this paper and the black
points are for nearby clouds, taken from 
\citet{Evans:2014}.
The black horizontal line is the mean value for the nearby clouds and
the red horizontal line is the mean value for the data in this paper.
The error bars at the far right are the standard deviations of $\log(\sfe)$.
}
\label{sfevsdense}
\end{figure*}

\begin{figure*}
\center
\includegraphics[scale=0.6, angle=0]{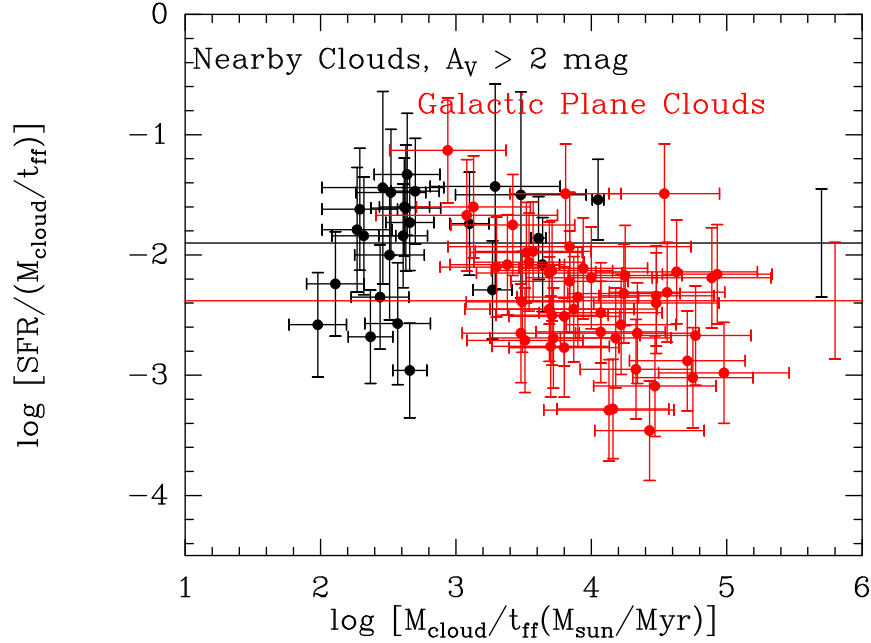}
\caption{
Plot of the logarithm of SFR divided by the cloud mass over
the free-fall time  vs. the logarithm of the cloud mass over the 
free-fall time. The red points are from this paper and the black
points are for nearby clouds, taken from 
\citet{Evans:2014}.
The black horizontal line is the mean value for the nearby clouds and
the red horizontal line is the mean value for the data in this paper.
The error bars at the far right of these lines 
are the standard deviation of the $\log(\sfr/(\mcloud/\tff))$ for the
two data sets.
}
\label{sfevsmtff}
\end{figure*}


\begin{deluxetable*}{llccl}
\tablecaption{Star Formation Efficiencies
\label{sfetab}}
\tablewidth{0pt}
\tablehead{
\colhead{Variable} &
\colhead{Log SFE(nearby)} & 
\colhead{Standard Deviation} & 
\colhead{Log SFE(GP)} &
\colhead{Standard Deviation}  
 }
\startdata
Cloud Mass & $-2.07$  & $0.83$  & $-2.91$ & $0.53$ \\
Dense Mass & $-1.61$  & $0.23$  &  $-1.74$ & $0.50$ \\
\mcloud/\tff & $-1.90$  & $0.45$  & $-2.38$ & $0.49$ \\
\enddata
\end{deluxetable*}

We plot the SFR per entity for the three main entities used to 
predict SFR in Figures \ref{sfevsmass} to \ref{sfevsmtff}.
The three entities are cloud mass (\mcloud), dense gas
mass (\md), and cloud mass per free-fall time
(\mcloud/\tff). In extragalactic parlance, the first two are 
``star formation efficiencies", or the reciprocal of the depletion time
\added{, measured in Myr}.
The last is unitless \added{and equal to \epsff}.
We will refer to these generically as the SFE.
The black points are from the nearby
clouds, while the red points are from this paper. 
The means and standard deviations (both in the log) are given in Table \ref{sfetab}.

The following facts are apparent.  
The standard deviations in the SFE for Galactic Plane clouds
are comparable (about 0.5 dex)
for all the relations
(the figures are plotted on the same scales so the
eyeball estimate of the scatter is meaningful).
\added{The mean of the SFE for the Galactic Plane clouds
is lower by 0.8 in the log than that for the nearby clouds when cloud mass is used, but very similar (lower by 0.13 in the log) when dense gas mass is used.}

The agreement of the SFE for the dense gas relation is striking because
the method to estimate the SFR for the Galactic Plane sample is very
different (MIR emission) from that used for the nearby clouds (YSO counts)
so disagreement could be expected. Indeed, we know that MIR emission
seriously underestimates the SFR for the nearby clouds, and we 
\replaced{could}{might}
expect an underestimate of 0.3 to 0.5 in the log for the Galactic Plane
sample, based on the discussion in \S 3.2.3. The agreement \deleted{here}
\added{for the dense gas}
encourages us that the Galactic Plane sources have SFR sufficiently high
that the MIR tracer is not greatly underestimating the SFR. 
\added{The alternative explanation, that the SFR and the dense gas
mass are both underestimated by the same amount, is implausible because
the mean density of the Galactic Plane clumps is less than that of the
dense clumps in the nearby clouds, as noted above. If anything, the masses
of gas as dense as those in the nearby clouds is over-estimated in the
Galactic Plane clouds.} 
\deleted{The mean values of SFE  are lower for the Galactic Plane sample, but least
so for the dense gas, for which they agree to within 0.13 in the log. }

The SFE for the Galactic Plane clouds is about \eten{-3} 
\added{Myr$^{-1}$}, which is
consistent with a depletion time of 1 Gyr, similar to that for galaxies
as a whole.
\added{
\citet{2011ApJ...729..133M} used WMAP data to find the most luminous \hii\
regions in the Galaxy and to compute star formation rates. 
His sample is clearly biased to the regions with the highest star formation rates. 
After connecting these \hii\ regions to molecular gas using catalogs
of GMCs, he computed star formation efficiences. For consistency, we used
our equation to recompute his star formation rates, but they are only slightly
lower than the rates he gave. 
His data are added in Figure \ref{murray} as the blue points; 
they have a mean
SFE similar to that of the nearby clouds, and they further extend the
dispersion among Galactic Plane clouds. 
We cannot do a full analysis with Murray's sources as no uncertainties
are given, the identification with molecular gas is not clearly described,
and there is no information on the dense gas. However, a simple mean value
of SFE including his sample and ours yields $\mean{\log {\rm SFE}} = -2.52
\pm 0.72$ where SFE is measured in Myr$^{-1}$ as usual, bringing the mean
for the Galactic Plane clouds closer to that for the nearby clouds.

\begin{figure*}
\center
\includegraphics[scale=0.6, angle=0]{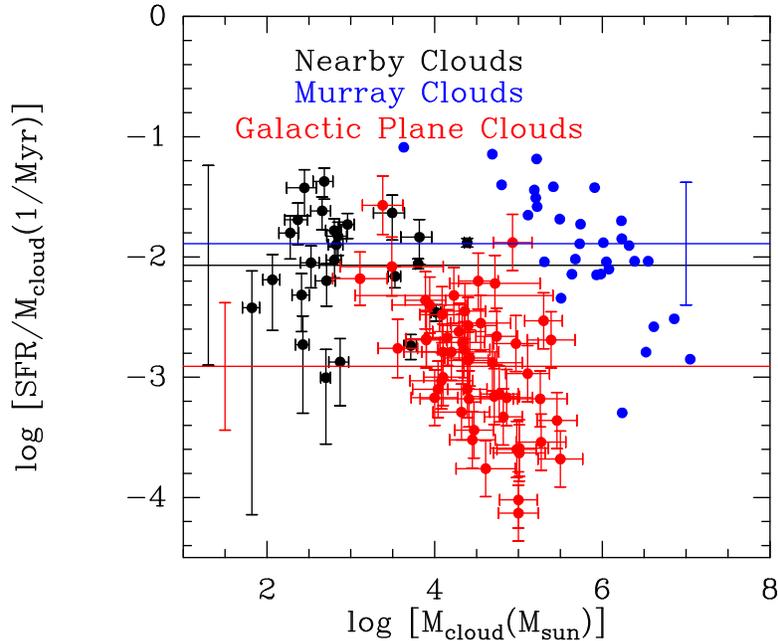}
\caption{
Plot of the logarithm of SFR per mass of molecular gas vs. the logarithm of
the mass of molecular gas. The red points and black points are as
in the previous figure. The blue points are from
\citet{2011ApJ...729..133M}.
The blue horizontal line is the mean value and error bar at the far right is the standard deviation for the data from 
\citet{2011ApJ...729..133M}.
}
\label{murray}
\end{figure*}
}

\deleted{This  also 
implies that the low SFE for the Galactic Plane clouds in Fig. \ref{sfevsmass} is real.}

\deleted{Far from being inefficient star formers, the nearby clouds
are efficient relative to the Galactic Plane clouds, although
this statement carries the caveat that the regions appear to be 
denser in the nearby clouds as discussed above.}

\replaced{A possible cause for the discrepancy of SFE found in nearby clouds and the Galactic Plane sample 
is radiative and mechanical feedback from massive stars. }
{The main feature that emerges from the data is a large dispersion
in SFR per cloud mass for both the nearby and the Galactic Plane clouds.
The mean SFE depends on sample selection. When the dense gas mass is
used to determine SFE, the dispersion is less and the mean values
for nearby and Galactic Plane clouds agree. Differences between the
nearby clouds and the Galactic Plane clouds may
be related to different feedback effects.}
Other than the Orion cloud, the local regions are primarily generating low mass stars \added{because the IMF is not fully sampled}. 
In such low mass star forming regions, protostellar outflows are the primary feedback process. 
Such outflows 
can perturb the cloud structure 
over scales up to several parsecs over a restricted volume set by the jet opening angle. 
However, their effect on the overall cloud structure is limited and 
incapable of suppressing the SFE.  Massive young stars provide a much stronger energy input to the 
cloud primarily through far-UV radiation fields that drive expanding \hii\ regions \citep{Matzner:2002, 
Dale:2014}. 
These processes can impact the SFE by photoionizing part of the cloud or simply 
modifying the conditions \replaced{that are}{to make them} less suitable for star formation. 
Some support for this might be found in the apparently higher mean
value of $\alpha$ in the Group 2 clouds (Table \ref{grouptab2})
but the dispersion is quite large. In addition, the mean values for
$\alpha$ for the nearby clouds and the Galactic Plane clouds are
indistinguishable.
\added{The most likely effect of the massive stars is to confuse
observational measures by removing or ionizing the molecular gas,
leading to Group 2 or 3 sources, configurations which are not seen in the nearby clouds.}

\section{Comparison to Extragalactic Results}\label{exgal}

Studies of other galaxies have explored the same questions
as have been addressed in this paper. In this section, we compare
those results to those in the Milky Way. We continue to use the SFE
to avoid the strong apparent correlation introduced by the fact that
both mass and SFR generally scale as size, hence distance, squared.
For the extragalactic studies, clouds are generally not resolved, so
we use \mmol\ to represent the aggregate mass in molecular gas, while
\mdense\ continues to represent the aggregate mass of dense gas.
\begin{figure*}
\center
\includegraphics[scale=0.6, angle=0]{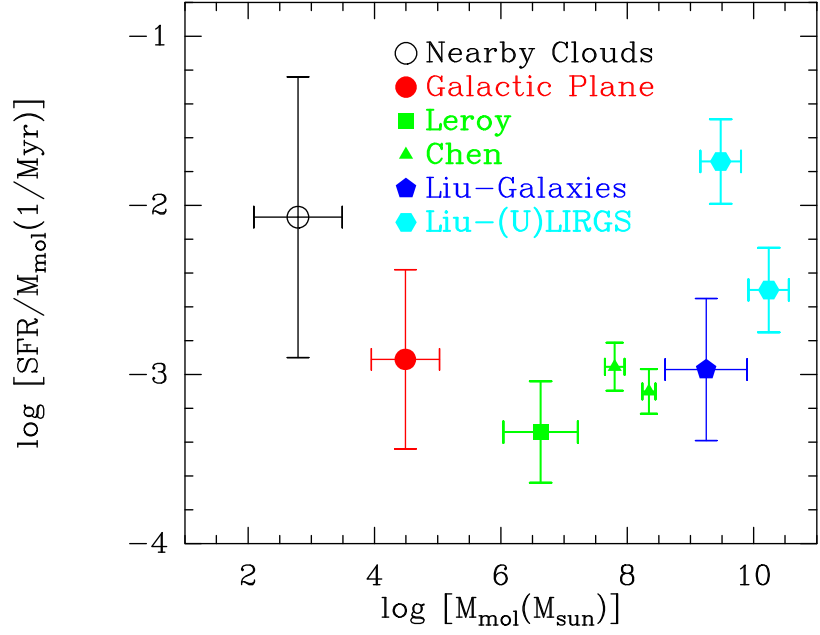}
\caption{
Plot of the logarithm of SFR per mass of molecular gas vs. the logarithm of
the mass of molecular gas, showing averages and standard deviations (in
the logs). The red filled circle is from this paper and the open black
circle is for the nearby clouds, taken from 
\citet{Evans:2014}.
Three green points are plotted for resolved studies
in nearby galaxies. The lowest point (square)  is from
\citet{2013AJ....146...19L}
who observed with resolutions from 0.2 to 1.4 kpc.
The other two points (triangles) are from 
\cite{2015ApJ...810..140C},
who observed M51 with 1 kpc resolution; the higher point is
for the outer galaxy and the lower point is for inner
($r < 1.66$ kpc) part of M51.
The blue point (pentagon) represents the normal galaxies 
while the cyan points (hexagons) represent the (U)LIRGs from
\citet{2015ApJ...805...31L}.
Two values are plotted,
the higher one using the conversion from CO to molecular gas of 
$\alpha_{\rm CO} = 0.8$ \msun (K \kms pc$^2$)$^{-1}$, 
while the lower one uses the same value as for the normal galaxies,
$\alpha_{\rm CO} = 4.6$ \msun (K \kms pc$^2$)$^{-1}$. 
}
\label{galmcloud}
\end{figure*}

\begin{figure*}
\center
\includegraphics[scale=0.6, angle=0]{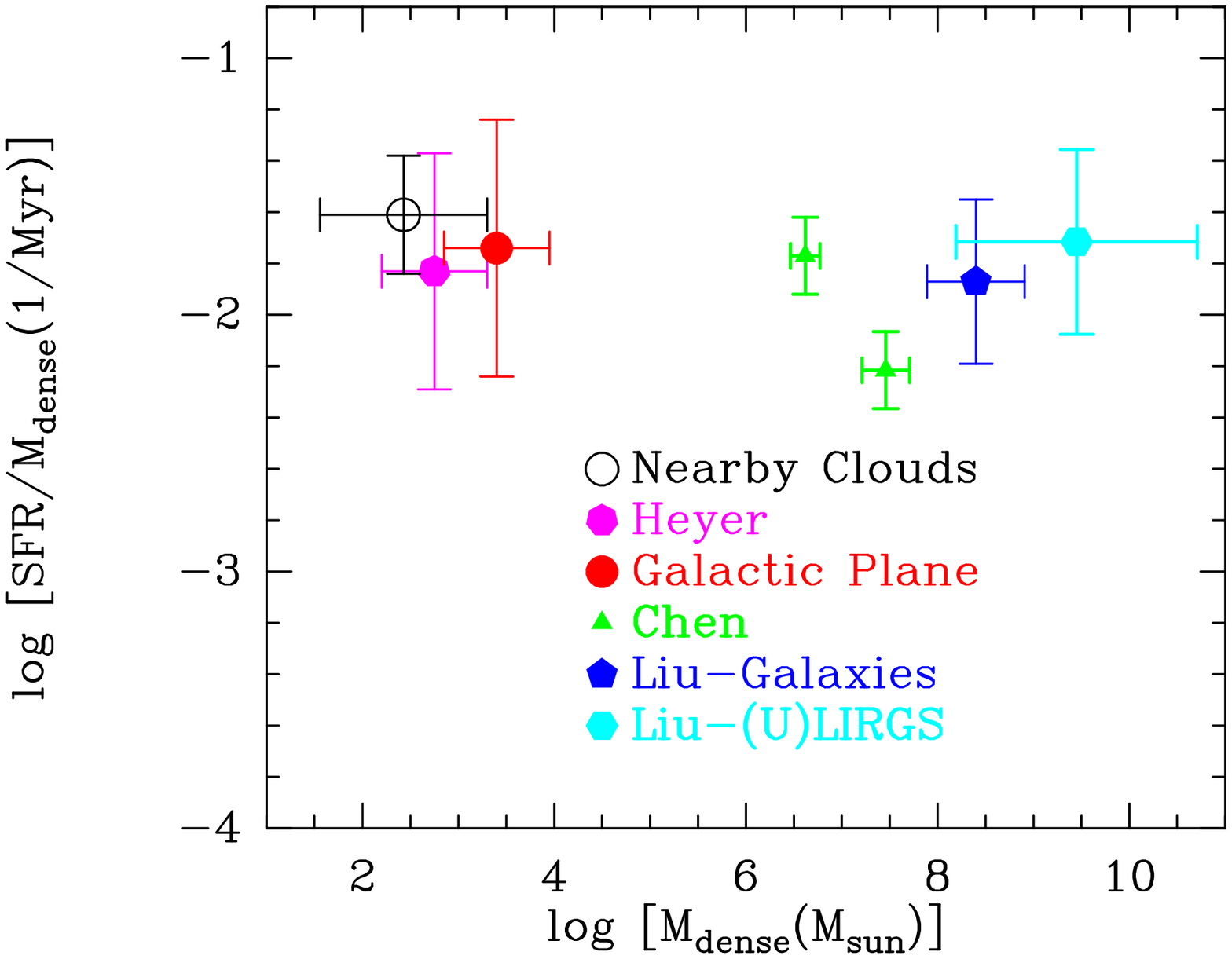}
\caption{
Plot of the logarithm of SFR per mass of dense gas vs. the logarithm of
the mass of dense gas, showing averages and standard deviations (in
the logs). The red point (filled circle) is from this paper and the black
point  (open circle) is for the nearby clouds, taken from 
\citet{Evans:2014}.
The magenta point (heptagon) comes from 
\citet{Heyer:2016}. 
The two green points (triangles) are from
\cite{2015ApJ...810..140C},
who observed M51 with 1 kpc resolution; the higher point is
for the outer galaxy and the lower point is for inner
($r < 1.66$ kpc) part of M51.
The  blue point (pentagon) represents the whole, normal galaxies 
while the cyan point (hexagon) represents the whole  (U)LIRGs, both from
\citet{2015ApJ...805...31L}.
}
\label{galmdense}
\end{figure*}

The first comparison is for overall molecular gas. Figure \ref{galmcloud}
plots the mean and standard deviation (in log space) of the
SFE for the nearby clouds and the Galactic Plane clouds, along
with data from other studies.
The green points show results from  spatially resolved studies of nearby
galaxies. The  lowest green point is from
\citet{2013AJ....146...19L},
who summarized the HERACLES data on 30 ``nearby" ($d < 22$ Mpc)
disk galaxies.  The SFE plotted for the galaxies is actually the log
of $1/\mean{\tdep}$, but calculation of \mean{SFE}\ from their
data tables produces a nearly identical result. The mass scale is
that of their resolution, obtained from the surface density and
resolution in their Table 1, converted to mass, converted to logs,
and averaged, resulting in a characteristic log(\mmol) of 
$6.63\pm0.59$. 
The other two green points are from the study of M51 with 1 kpc
resolution as presented by
\citet{2015ApJ...810..140C};
the highest green point is for the outer parts of M51, while
the lower one at higher masses is from the inner 1.66 kpc.
Finally, whole galaxies can be plotted, based on a sample of 115 disk
galaxies in 
\citet{2015ApJ...805...31L}. 
The normal galaxies are plotted in blue and the 
\ulirg s are plotted in cyan. The higher cyan point uses the
\ulirg\  conversion from CO to molecular gas while the lower
point uses the same conversion as used for the normal galaxies.
The plot shows a decline in SFE from the nearby
clouds to the Galactic Plane clouds to the extragalactic regions.
The standard deviation in the SFE also decreases, presumably due
to averaging over more and more star formation regions
(cf \citealt{Kruijssen:2014,Vutisalchavakul:2014}).
A rise in SFE is seen for the \ulirg s, especially if the lower,
\added{(U)LIRG,} value
for the conversion factor is used, for which the SFE exceeds even
the value in the nearby clouds.

The same plot can be made for the dense gas, again using
averages and standard deviations in logs, using masses based
on HCN emission and 
\begin{equation}
\mdense(\msun)  = 10 L_{\rm HCN} 
\end{equation}
where $L_{\rm HCN}$ is the observed luminosity of HCN \jj10\ emission
in K \kms\ pc$^2$
\citep{2015ApJ...805...31L}.
The \ulirg s  in \citet{2015ApJ...805...31L} are again plotted in cyan,
but only one value of conversion from HCN emission to dense gas
mass was used by \citet{2015ApJ...805...31L}.
Observations of HCN in M51 with a resolution of 1 kpc bridge the gap
between the Milky Way clouds and whole galaxies. The green points
are based on data in
\citet{2015ApJ...810..140C};
the higher point is the average for the outer part, while the
lower point is for the inner part of M51.
A magenta point comes from the study of Milky Way clouds by 
\citet{Heyer:2016}
who derived star formation rates from mid-infrared luminosities of Class I protostars in a sample of dense 
clumps identified in the Atlasgal survey of submillimeter dust emission
\citep{Heyer:2016}.

Figure \ref{galmdense} shows that the SFE measured for dense gas
varies little from local clouds to whole galaxies.
These results are very consistent with those found by 
\citet{2015ApJ...805...31L},
but extend them down to the scales of individual clouds in the Galaxy,
and even down to the scales of nearby clouds, where the methods of measuring
both SFR and \mdense\ are quite different. The basic conclusion is that
the  mass of dense gas is the most stable predictor of SFE across
seven orders of magnitude in dense gas mass.
The grand average of the data points in Figure \ref{galmcloud} yields
$\mean{{\rm log\ SFE(Myr^{-1})}} =  -2.83\pm0.42$ 
\explain{corrected to add the units and that the mean is in the log}
if the \ulirg s with the Milky Way
value for $\alpha_{\rm CO}$ are included, or $-2.73\pm 0.59$ if the
point using the \ulirg\ value for $\alpha_{\rm CO}$ is included.
In contrast, the grand average for the dense gas data in Figure
\ref{galmdense} is $\mean{{\rm log\ SFE(Myr^{-1})}} = -1.82\pm 0.19$, 
with a standard deviation
one-third that when all molecular gas is used.

\replaced{However,}{Despite the remarkable consistency shown in
Figure \ref{galmdense},} 
\mdense, at least as measured by HCN, is not perfect;
studies of our Galactic Center and nearby galaxies
using HCN \jj10\ as the tracer of dense gas 
\citep{Usero:2015,2015ApJ...810..140C,2013MNRAS.429..987L}
have shown that the SFE for dense gas also depends on environment. 
\added{This dependence can be seen in Figure \ref{galmdense} where the
lowest point is for the inner part of M51. Recent studies show that this
decrease in SFE is a function of various galactic properties, such
as molecular fraction \citep{2016ApJ...822L..26B}. The most plausible
explanation for lower SFE for gas probed by HCN emission in regions
of high density is that the criterion for rapid star formation changes.
\citet{2014MNRAS.440.3370K} provide an exhaustive examination of
both global and local mechanisms to explain the low SFE for gas 
probed by \ammonia\ emission in the central
molecular zone of our Galaxy. They favor episodic star formation among
global mechanisms and a much higher threshold density for star formation
among local mechanisms. The higher threshold is caused by the greatly increased
turbulence in the central molecular zone, pushing the density threshold
to values as high as $n \approx \eten7$ \cmv, far above the density needed
to produce strong emission from HCN \jj10\ 
\citep{1999ARA&A..37..311E,2015PASP..127..299S}. 
In these environments, probes of higher densities, such as higher $J$ transitions of HCN, may
be more diagnostic of the mass of gas above the threshold density.}
Further studies of central regions of galaxies, including our own,
are needed to refine the criteria for star formation.

\added{
There are many reasons why measures of SFE in the range of regions
plotted in figures \ref{galmcloud} and \ref{galmdense} might differ.
The methods used to determine cloud mass and dense gas mass differ and
the meaning of ``dense'' is not the same for all. Extragalactic studies,
especially whole galaxy observations, average over a huge range of
physical conditions and sample regions at different stages of
evolution, while studies of individual clouds are snapshots \citep{Kruijssen:2014}. The fact that the differences in SFE
across this range are so small when measured against dense gas is
an important clue for our understanding of what controls star formation.
The consistency suggests that simulations of star formation and galaxy
evolution that require higher densities to initiate star formation
are on the right track.
A picture in which most molecular clouds are unbound and only small, dense parts of the clouds are sites of star formation
\citep{2011MNRAS.413.2935D}
or a picture in which dense clump formation \replaced{is delayed}{from more
diffuse molecular gas is a continuous process \citep{2013ApJ...773...48B}} can potentially explain our results. 
 Pictures in which a high density
{\it contrast} is needed for star formation 
may be able to incorporate the lower SFE for dense gas in galaxy centers
\citep{2016ApJ...822L..26B}.
A combination of these ideas may produce a more unified picture.
Future work on other galaxies with high
spatial resolution, using dense gas tracers, will test these relations
further.
}

\section{Summary}

We compiled a sample of Galactic \gmc s that are associated with \hii\
regions and estimated their properties
and SFR. The analysis of \gmc s, \hii\ regions, 
and SF tracers (both radio continuum and MIR emission)  shows
different degrees of associations between molecular gas and
star formation. We classified the \gmc s into different groups:
 \gmc s with embedded \hii\ regions, \gmc s
with overlapping \hii\ regions, and \gmc s with separated \hii\ regions. 
We did not use the last group because association between molecular gas
and star formation was too uncertain.

The  sample was used to test relations between SFR and properties
of \gmc s. We tested \replaced{five}{four} different models of star formation.
No significant correlation was found between 
\epsff\ and $\tff/\tdyn$.
Significant correlations exist between SFR and \mcloud, \md,
and $\mcloud/\tff$.
The relation between SFR and \md\ is consistent with linear, while
the other two are significantly \replaced{sub-linear}{non-linear}, unlike extragalactic
relations or the theoretical model by \citet{Krumholz:2012}
for $\mcloud/\tff$.

Combining the data 
\added{on Galactic Plane clouds presented in this paper}
 with that on 
nearby clouds shows that the star
formation efficiency of the nearby clouds is higher when efficiency
is measured versus \mcloud\ or \mcloud/\tff. The efficiency per
mass of dense gas is very similar for the nearby clouds and the Galactic
plane clouds.

Adding extragalactic studies, we can extend the range of relevant
mass scales over 7 orders of magnitude. The star formation efficiency
for dense gas shows remarkable stability over this range, varying
over a factor of 4, while that
for total molecular gas varies by a factor of 40.
\added{The standard deviation in the log of the SFE(Myr$^{-1}$)
decreases by about a factor
of 3 to a value of 0.19 when dense gas mass, rather than molecular mass, is used.}

\medskip

\added{We thank the anonymous referee for a careful reading and excellent
suggestions which have improved the paper. We also thank C. McKee,
C. Federrath, G. Parmentier, M. Fall, and B. Ochsendorf for comments.}
We are grateful to the BGPS team for sharing ideas and information
over many years. We particularly thank B. Svoboda for calculations
to characterize the extinction threshold for the BGPS sample.
H. Chen kindly provided data on galaxies.
This work was supported by NSF
grant AST-1109116 to the University of Texas at Austin.
MH acknowledges support from NASA ADAP grant NNX13AF08G 
to the University of Massachusetts.


\appendix{}

\section{Error Analysis} \label{errors}

The uncertainties are in general propagated from those arising from
the fundamental measurements.
Distance errors are not included in the values in the tables such
as cloud mass, dense gas mass, and star formation rate because
distance cancels out in some quantities, such as star formation 
efficiency. However, where a star formation rate is plotted versus
something like a mass, or where correlation fits are done, a distance
error has been incorporated by adding it in quadrature to the uncertainties
in SFR or mass. 
\added{We use \sfrp\ and \mclp\ to represent the \sfr\ and \mcloud\
without the distance factors, so that $\sfr = \sfrp D^2$ and 
$\mcloud = \mclp D^2$.
}
When $\sigma(\sfrp)$ or $\sigma(\mclp)$ 
appear in the equations below, they do not include the distance
uncertainties, represented by $\sigma(D)$, but the uncertainty
in \sfrp\ includes a nominal 10\% uncertainty and the cloud mass
estimate includes uncertainties
from excitation and abundance as follows:
\begin{equation}
\sigma(\mclp) = \mclp [ (\sigma(N(^{13}CO))/N(^{13}CO))^2 + 
(\sigma(X(\coo))/X(\coo))^2]^{0.5} ,
\end{equation}
where $\sigma(N(^{13}CO))/N(^{13}CO) = 0.27$, $\sigma(X(\coo))/X(\coo) = 0.30$,
and $X(\coo) = \hh/\coo$
(\S 3.3.1).

For more complex variables, the dependence
on distance is based on analysis of how the final quantity depends
on the fundamental variables.
The quantity $\mcloud/\tff$ has the following dependency:


\added{
\begin{equation}
\mcloud/\tff \propto \mcloud \times \sqrt{\rho}\
\propto \mcloud \times \sqrt{\frac{\mcloud}{\rcloud^3}}\
\propto \frac{\mcloud^{1.5}}{\rcloud^{1.5}}
\propto \frac{\mclp^{1.5} D^{3.0}}{\thetacl^{1.5} D^{1.5}}
\propto \frac{\mclp^{1.5} D^{1.5}}{\thetacl^{1.5}}
\end{equation}
since the uncertainty in cloud size is proportional to distance with
fixed angular resolution.  Then,
\begin{equation}
\sigma(\mcloud/\tff) = (\mcloud/\tff) 
[(1.5 \sigma(\mclp)/\mclp)^2  + (1.5 \sigma(D)/D)^2]^{0.5}
\end{equation}
because uncertainties in \thetacl\ are negligible compared to those
in distance.
}


Because $\sfr = \sfrp D^2$, the efficiency quantity, 
\added{
\begin{equation}
\epsff = \sfr/(\mcloud/\tff) \propto \frac{\sfrp\ D^{2}}{\mclp^{1.5} D^{1.5}}
\propto \sfrp\ \mclp^{-1.5} D^{0.5}
\end{equation}
and
\begin{equation}
\sigma(\epsff) = \epsff [(\sigma(\sfrp)/\sfrp)^2 + (0.5\sigma(D)/D)^2
+ (1.5 \sigma(\mclp)/\mclp)^2]^{0.5}
\end{equation}
}

The quantity \tff/\tdyn\ is
\begin{equation}
\tff/\tdyn = 0.86 \sqrt{\alpha}
\end{equation}
and
\begin{equation}
\sigma(\tff/\tdyn)= (\tff/\tdyn) |{0.5 \sigma(\alpha)/\alpha}|
\end{equation}
where
\begin{equation}
\sigma(\alpha) = \alpha [ (2\sigma(v)/v)^2 + (\sigma(\mclp)/\mclp)^2
+ (\sigma(D)/D)^2]^{0.5}
\end{equation}
because
\begin{equation}
\alpha = \mvir/\mcloud \propto \frac{\delta v^2 D}{\mclp D^2}
\propto \delta v^2 D^{-1}
\end{equation}

\bibliographystyle{apj}
\bibliography{cite} 

\end{document}